\documentclass[aps,pre,superscriptaddress]{revtex4}
\usepackage[dvips]{graphics}
\usepackage{graphicx}
\usepackage{amsfonts}
\usepackage{amssymb}
\usepackage{amsmath}
\usepackage{subfigure}

\begin{document}

\title{Collisionally inhomogeneous Bose-Einstein condensates in double-well potentials
}

\author{C. Wang }
\affiliation{Department of Mathematics and Statistics, University of Massachusetts,
Amherst MA 01003-4515, USA}

\author{P.G.\ Kevrekidis }
\affiliation{Department of Mathematics and Statistics, University of Massachusetts,
Amherst MA 01003-4515, USA}

\author{N. Whitaker}
\affiliation{Department of Mathematics and Statistics, University of Massachusetts,
Amherst MA 01003-4515, USA}

\author{D.\ J.\ Frantzeskakis}
\affiliation{Department of Physics, University of Athens, Panepistimiopolis, Zografos, Athens 157 84, Greece}

\author{P.\ Schmelcher}
\affiliation{Theoretische Chemie, Physikalisch-Chemisches Institut, Im Neuenheimer Feld 229,
Universit\"at Heidelberg, 69120 Heidelberg, Germany}
\affiliation{Physikalisches Institut, Universit\"at Heidelberg, Philosophenweg 12, 69120 Heidelberg, Germany}

\author{S.\ Middelkamp}
\affiliation{Theoretische Chemie, Physikalisch-Chemisches Institut, Im Neuenheimer Feld 229,
Universit\"at Heidelberg, 69120 Heidelberg, Germany}

\begin{abstract}
In this 
work, we consider quasi-one-dimensional 
Bose-Einstein condensates (BECs), with spatially varying collisional interactions, trapped in 
double well potentials. In particular, we study a setup in which such a 
``collisionally inhomogeneous'' BEC has the same (attractive-attractive or repulsive-repulsive) or different 
(attractive-repulsive) type of interparticle interactions. 
Our analysis is based on the continuation of 
the symmetric ground state and anti-symmetric first excited state of the non-interacting 
(linear) limit into their nonlinear counterparts.
The collisional inhomogeneity produces a saddle-node bifurcation
scenario between two additional solution branches; as the inhomogeneity 
becomes stronger, the turning point of the
saddle-node tends to infinity and eventually only the two 
original branches remain present, 
which is completely different from the standard double-well phenomenology.
Finally, one of these branches changes its monotonicity as a function of
the chemical potential, a feature especially prominent, when the 
sign of the nonlinearity changes between the two wells. Our 
theoretical predictions, 
are in excellent agreement with the numerical results.
\end{abstract}

\maketitle

\section{Introduction} 

The remarkable progress in the experimental and theoretical
studies of Bose-Einstein condensates (BECs) \cite{book1,book2} over the past two decades 
has sparked an intense study of the coherent nonlinear structures 
that arise in this setting. Such 
structures are 
rather naturally expected to emerge due to a well-established
mean-field description of BECs by the Gross-Pitaevskii (GP) equation, 
namely, a nonlinear Schr{\"o}dinger (NLS) equation, in which the nonlinearity is 
introduced by the interatomic interactions.  
This feature has inspired many relevant theoretical and experimental studies 
devoted to macroscopic nonlinear matter waves, such as bright 
matter-wave solitons in attractive BECs \cite{expb1,expb2,expb3}, 
as well as dark \cite{dark1,dark2,dark3,dark4} and gap \cite{gap}
matter wave solitons in repulsive BECs (see also the recent review \cite{ourbook}). One of the important 
elements of the versatility of this atomic physics setting 
is the existence of diverse types of external potentials in 
the GP equation accounting for the electric, magnetic, optical or combined confinement 
of ultracold dilute alkali vapors that constitute the BEC. Some of the most typical forms of such trapping potentials 
include a parabolic and a spatially periodic one (created by the
interference of counter-propagating laser beams, 
the so-called optical lattice). The NLS equations with similar potentials are also relevant 
in the context of nonlinear optics, where they model the evolution of 
optical beams in graded-index waveguides and periodic waveguiding arrays
\cite{kivshar,reviewsopt}. 

One interesting type of potential that has drawn a considerable
amount of attention 
is the 
double-well potential.
This may originate, for instance, from the combination of a parabolic
with a periodic potential. Its experimental realization
was featured in Ref. \cite{markus1}, where 
a variety of interesting phenomena were studied; 
these include Josephson oscillations and
tunneling for a small number of atoms, or macroscopic quantum
self-trapping and an asymmetric partition of the atoms between
the wells for sufficiently large numbers of atoms. 
In parallel to these experimental findings, numerous theoretical
insights on this topic have emerged 
\cite{smerzi,kiv2,mahmud,bam,Bergeman_2mode,infeld,todd,theo,carr}.
These concerned finite-mode reductions, analytical results for
specially designed shapes of the potential, quantum depletion effects, and
other theoretical aspects. 
Interestingly, 
double-well potentials have also been studied in applications arising in 
the context of nonlinear optics, including 
twin-core self-guided laser beams in Kerr media \cite{HaeltermannPRL02}, optically
induced dual-core waveguiding structures in a photorefractive crystal \cite{zhigang}, 
trapped light beams in a structured annular core of an optical fiber \cite{Longhi}, and so on. 
It is relevant to point out here that double well settings have been
examined not only in one-component systems, but also in multi-component
cases. In particular, a recent study motivated by two-component
BECs can be found in \cite{chenyu_two}, while similar attempts have
been made in the context of the so-called spinor BECs (where it is
possible to have three, and even five components) in the works of
\cite{liyou1,liyou2}. These works examined not only finite-mode
reductions of the multi-component case, but also phenomena beyond
the level of the mean-field description such as quantum entanglement
and spin-squeezing properties.

On the other hand, nonlinear matter-waves have 
not only been studied in a variety 
of external potentials, but also in the presence of temporally or spatially varying 
external fields manipulating the interatomic interactions.
Indeed, the s-wave scattering length (which characterizes the nonlinearity coefficient in the GP equation) 
can be adjusted experimentally by employing 
either magnetic \cite{Koehler,feshbachNa} or optical Feshbach 
resonances \cite{ofr} in a very broad range. This flexibility of 
manipulation of the interatomic interactions has motivated
a significant number of studies both on the theoretical and on
the experimental front; in particular, on the experimental side,
the formation of bright matter-wave solitons and soliton trains for
$^{7}$Li \cite{expb1,expb2} and $^{85}$Rb \cite{expb3} atoms
used a tuning of the interatomic interactions from repulsive to attractive. 
Also, this type of manipulations was instrumental
in achieving the formation of molecular condensates \cite{molecule}, 
and the probing of the BEC-BCS crossover \cite{becbcs}.
On the other hand,  theoretical studies have predicted that a time-dependent 
modulation of the scattering length can be used to  
stabilize attractive higher-dimensional BECs against collapse \cite{FRM1}, or
to create robust matter-wave breathers in lower-dimensional BECs \cite{FRM2}.
While the above studies focused on temporal variations of the 
interaction strength, more recently spatial variations of the nonlinearity
have come to be of interest in the so-called ``collisionally inhomogeneous''
environments. These have been found to lead to a variety of 
interesting developments including (but not limited to) 
adiabatic compression of matter-waves \cite{our1,fka}, 
Bloch oscillations of matter-wave solitons \cite{our1},  
atomic soliton emission and atom lasers \cite{vpg12}, 
enhancement of transmittivity of matter-waves through barriers \cite{our2,fka2}, 
dynamical trapping of matter-wave solitons \cite{our2}, 
stable condensates exhibiting both attractive and repulsive interatomic interactions \cite{chin}, the
delocalization transition of matter waves~\cite{LocDeloc}, and so on.
Many different types of spatial variations of the nonlinearity have 
been considered, including linear \cite{our1,our2}, parabolic \cite{yiota}, random \cite{vpg14}, 
periodic \cite{vpg16,LocDeloc,BludKon}, and localized (step-like) \cite{vpg12,vpg17,vpg_new} 
ones. Furthermore, a number of detailed mathematical
studies \cite{key-2,key-4,vprl} have appeared, addressing aspects
such as the effect of a ``nonlinear lattice potential" (i.e., a spatially 
periodic nonlinearity) on the stability
of 
matter-wave solitons, and the interplay between drift and diffraction/blow-up 
instabilities. 
More recently, the interplay of nonlinear and linear potentials has been 
examined in both continuum \cite{ckrtj} and discrete \cite{blud_pre} settings.

The aim of the present work is, in fact, to combine these
two interesting settings, namely the double-well 
potential and a collisionally inhomogeneous (i.e., spatially dependent) nonlinearity.
As, arguably, one of the simplest forms of this combination, we 
select a coefficient of the nonlinearity which is piecewise 
constant, in line with the suggestions of \cite{vpg12,vpg_new,rodrigues}
and can vary between two (smoothly connected) pieces with the same
sign, or even two pieces with opposite signs. 
The latter form of a spatially dependent nonlinearity appears to be
one that should be straightforwardly experimentally realizable \cite{krueger},
through the use of magnetic field gradients of moderate size for atom chips
(see also the relevant discussion in \cite{rodrigues}).
The phenomenology
that we observe in this setting appears to be {\it remarkably
different} from that of the standard double-well. In particular,
as we detailed in an earlier relevant work 
\cite{theo}, the linear states of the underlying potential can
be continued into nonlinear ones 
in the presence of the pertinent cubic nonlinearity. 
As prototypical examples of these linear states 
one can consider the symmetric ground state 
and the anti-symmetric first excited state of the double-well potential. 
A remarkable feature, however,
which takes place for sufficiently strong nonlinearity (to the symmetric 
state 
for attractive interactions 
and to the anti-symmetric state 
for repulsive interactions) 
is a symmetry-breaking bifurcation of the pitchfork type. 
This 
bifurcation generates two new asymmetric
solutions (which exist for sufficiently large nonlinearity).
What we find here is that even a weak 
inhomogeneity in the collisional properties of the condensate 
changes the nature of this bifurcation from a pitchfork to a 
saddle-node one. This may be anticipated given the ``non-parametrically
robust'' nature of the pitchfork bifurcation which will yield
similar changes in the presence of different asymmetries 
(see also \cite{theo}).
In addition to that, increasing the inhomogeneity strength 
shifts (eventually to infinity) the turning point of the saddle-node
bifurcation. Thus, for sufficiently large variation of the nonlinear 
coefficient between the two wells, only two nonlinear states emanate 
from the ground and first excited states of the linear problem. 
Furthermore, for one of these states, the monotonicity of the 
number of particles 
changes as a function of its nonlinear eigenvalue parameter (i.e., the chemical potential).
This is a rather unusual feature that leads to a bifurcation 
diagram entirely different from those of the homogeneous interatomic 
interactions 
(i.e., homogeneous nonlinearity) limit of either the one-component
\cite{theo} or of the multi-component system 
(see e.g., \cite{chenyu_two}).\footnote{We note in passing that 
for the multi-component system the diagram is also
substantially different (in that case, more involved) 
than that of the single component one, due to
the presence of ``mixed states'', mixing the symmetric and anti-symmetric
nonlinear states of {\it each} of the two components 
(and bifurcations thereof).}
These traits  are captured accurately, as we will show below, 
by a two-mode, Galerkin-type approximation \cite{todd,theo}
applied to the present collisionally inhomogeneous double well setting. 

Our presentation is structured as follows. The model and the 
semi-analytical predictions regarding its stationary solutions
are presented in section II. Our corresponding directly 
numerical findings and their comparison 
with the analysis are given in section III. 
In section IV, we present an alternative theoretical viewpoint more akin to
the model dynamics and compare its results to the full results
of the partial differential equation.
Finally, in section V, 
we present our summary and conclusions.

\section{The model and its semi-analytical consideration}

%

The dynamics of a quasi one-dimensional (1D) condensate, oriented along the $x$-axis, 
can be described by the following GP equation (see, e.g., Ch. 1 in \cite{ourbook} and \cite{gpe1d}), 
\begin{equation}
   i \hbar \partial_{t} \Psi = \left(-\frac{\hbar^2}{2m} + V(x) + g|\Psi|^{2} -\mu \right)\Psi,
\label{gpe}
\end{equation}   
where $\Psi(x,t)$ is the mean-field order parameter, $m$ is the atomic mass, and $\mu$ is
the chemical potential of the effectively 1D system. The nonlinear coefficient
$g$ arises from the interatomic interactions and has an effective 1D form, namely 
$g=2\hbar \omega_{\perp}a$, where $\omega_{\perp}$ is the transverse confining frequency and $a$ 
is the three-dimensional (3D) s-wave scattering length (the cases $a>0$ or $a<0$ correspond, respectively, 
to repulsive or attractive interatomic interactions). The external potential $V(x)$ in the 
GP Eq. (\ref{gpe}) consists of a regular harmonic trap and a  
repulsive potential localized at the harmonic trap center, namely,  
\begin{equation}
V(x) = \frac{1}{2}m \omega_{x}^2 x^2 + V_{0}\mathrm{sech}^{2}\left(\frac{x}{w_{\rm B}}\right),
\label{pot}
\end{equation}   
where $\omega_{x}$ is the longitudinal confining frequency, while $V_0$ and $w_{\rm B}$ are, respectively, 
the strength and width of the localized potential; the latter is in fact a barrier potential that 
may be created by a blue-detuned laser beam, repelling the atoms in the condensate. 
It is clear that the combination of the harmonic trap and the barrier potential is in fact a double-well potential.

Moreover, we assume that the collisional properties of the condensate are spatially inhomogeneous, 
i.e., $a=a(x)$, with the function $a(x)$ taking different, but smoothly connected, values in the two wells 
of the external potential. In particular, we consider that an external magnetic or optical field (see below) 
modifies the scattering length of the condensate as follows, 
\begin{equation}
a(x) = a_0 +a_1 \mathrm{tanh}\left(\frac{x}{W}\right),
\label{aofx}
\end{equation}   
where $a_0$ and $a_1$ are constant values of the condensate's scattering length in the absence 
and in the presence of the external field, respectively, which are smoothly connected through the 
$\mathrm{tanh}$ function (here, $W$ is the spatial length scale on which 
the transition between the two values $a_0$ and $a_1$ takes place). 
Apparently, far away from the harmonic trap center at $x=0$ (or, in other words, far away from the barrier), the scattering 
length takes the values $a=a_0 \pm a_1$, for $x \rightarrow \pm \infty$ respectively. Such an inhomogeneity of the scattering 
length may be realized in practice upon employing a bias homogeneous field and imposing a steep localized gradient on top of it. 
In such a case, in a quasi 1D configuration (as is the case under consideration) this will lead 
to a constant scattering length with value $a=a_1$ in the left well, which is followed by a
localized change of $a$, finally ending up with a second value $a=a_2$ in the right well. 
Notice that in our model we choose the function $\mathrm{tanh}$ to analytically describe the transition between these 
different constant values of $a$ since there are no ideal steps, but rather close approximations to it. 
Also, we naturally assume that we are relatively close to a Feshbach resonance, in order to easily manipulate the
scattering length with the external field.

Next, measuring time in units of the inverse transverse trapping frequency $\omega_{\perp}^{-1}$,  
length in units of the transverse oscillator length $l_{\perp} \equiv \sqrt{\hbar/m \omega_{\perp}}$, energy in 
units of $\hbar \omega_{\perp}$, and introducing the normalized wavefunction $u(x,t) = (2|a_{r}|)^{1/2}\Psi(x,t)$ 
(where $a_r$ is a reference value of the scattering length), we reduce the GP Eq. (\ref{gpe}) to the following dimensionless form:
\begin{equation}
   i\partial_{t}u = Hu + \Gamma(x)|u|^{2}u - \mu u.
\label{eq1}
\end{equation}   
In the above equation, 
\begin{equation}
   H \equiv -\frac{1}{2}\partial _{x}^{2} + V(x)
\label{eq2}
\end{equation}
is the ``single-particle'' operator in the normalized external potential 
\begin{equation}
V(x)=\frac{1}{2}\Omega ^{2}x^{2}+V_{0}\,\mathrm{sech}^{2}\left(\frac{x}{w}\right),
\label{eq3}
\end{equation}
with $\Omega \equiv \omega_x / \omega_{\perp}$ being the normalized harmonic trap strength and 
$w = w_{\rm B}/l_{\perp}$. The nonlinearity coefficient $\Gamma(x)$ in Eq. (\ref{eq1}) 
is given by 
\begin{equation}
\Gamma(x)=\alpha+\beta\,\mathrm{tanh}(bx), 
\label{Gamma}
\end{equation}
where $\alpha= a_1/|a_r|$, $\beta = a_2/|a_r|$, and $b=l_{\perp}/W$. 
Apparently, the cases $\Gamma>0$ or $\Gamma<0$ correspond to repulsive or 
attractive interactions, respectively. We finally mention that the number of 
particles in the condensate $\cal{N}$ is now defined as $\cal{N}$$=(l_{\perp}/2|a_r|)N$, 
where 
\begin{equation}
N=\int_{-\infty}^{+\infty} |u(x,t)|^2 dx
\label{N}
\end{equation}
is an integral of motion (normalized mumber of particles) 
of the GP Eq. (\ref{eq1}). Also, in our analysis (and particularly in the 
typical numerical results that will be presented below) 
we will assume fixed parameter values $\Omega=0.1$, $V_{0}=1$, $w=0.5$ and $b=1$. 

\begin{figure}[tbhp!]
     \centering
           \includegraphics[width=.4\textwidth]{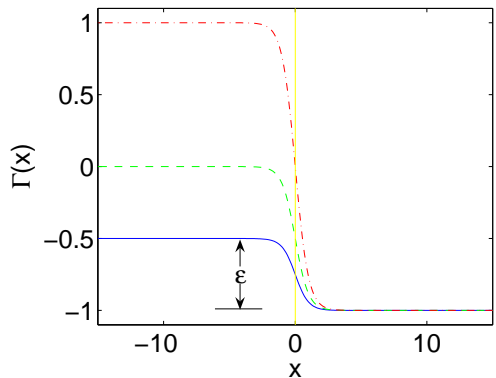}
     \caption{
The nonlinearity coefficient $\Gamma(x)=\alpha+\beta\,\mathrm{tanh}(bx)$ with parameters 
$\alpha+\beta=-1$, $\alpha-\beta=-1+\varepsilon$, and $b=1$. The
figure displays some examples of the shape of $\Gamma(x)$ for different values of $\varepsilon$ (i.e., the inhomogeneity parameter): 
$\varepsilon=0.5$ (blue solid line), $\varepsilon=1$ (green dashed line) and $\varepsilon=2$ (red dashed-dotted line.)}
      \label{fig1}
\end{figure}

\begin{figure}[tbhp!]
     \centering
           \includegraphics[width=.4\textwidth]{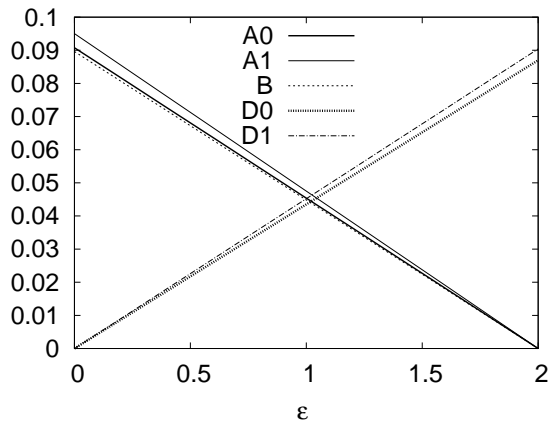}
     \caption{The dependence of the overlap integrals  
$A_0$, $A_1$, $B$, $D_0$ and $D_1$ on $\epsilon$ (for our choice of the
rest of the parameters of the potential and of $\Gamma(x)$).}
      \label{fig1a}
\end{figure}

In the non-interacting limit [i.e., for $\Gamma(x)=0$], the spectrum of the underlying linear Schr{\"o}dinger 
equation consists of a ground state, $u_{0}(x)$, and excited states, $u_{l}(x)$ ($l \ge 1$). 
Our analytical 
investigation of 
the nonlinear problem (at the stationary level)
in the weakly nonlinear regime consists of applying a Galerkin-type, two-mode approximation 
to Eq. (\ref{eq1}). In particular, we assume that 
the wavefunction $u(x,t)$ can be decomposed as 
\begin{equation}
u(x,t)=c_{0}(t)u_{0}(x)+c_{1}(t)u_{1}(x), 
\label{eq4}
\end{equation}
where $c_{0}(t)$ and $c_{1}(t)$ are unknown time-dependent complex prefactors. 
Substituting Eq. (\ref{eq4}) into Eq. (\ref{eq1}) and projecting the resulting equation 
to eigenfunctions $u_{0}$ and $u_{1}$, we 
obtain the following equations:
\begin{eqnarray}
i\dot{c_{0}}&=&(\omega _{0}-\mu)c_{0}+ A_{0}|c_{0}|^{2}c_{0}+ D_{0}(2|c_{0}|^2c_{1}+ c_{0}^2 c_{1}^{*})+ B(2|c_{1}|^{2}c_{0}+c_{1}^{2}c_{0}^{*})+ D_{1}|c_{1}|^{2}c_{1}),
\label{eq5}
\\
i\dot{c_{1}}&=&(\omega _{1}-\mu)c_{1}+ A_{1}|c_{1}|^{2}c_{1}+ D_{1}(2|c_{1}|^2c_{0}+ c_{1}^2 c_{0}^{*})+ B(2|c_{0}|^{2}c_{1}+c_{0}^{2}c_{1}^{*})+ D_{0}|c_{0}|^{2}c_{0}).
\label{eq6}
\end{eqnarray}
In these equations, dots denote time derivatives, stars denote complex conjugates, and $\omega_{0,1}$ are the eigenvalues 
corresponding to the eigenstates $u_{0,1}$; in the numerical examples
presented herein, $\omega_{0}$ and $\omega_{1}$  can be numerically
found to be $0.133$ and $0.156$ respectively. 
We should note in passing here that although $\omega_{0,1}$ and $u_{0,1}$
are numerically obtained through a simple eigensolver of the underlying
linear Schr{\"o}dinger problem, it is also, in principle, possible to 
develop a perturbative approach towards obtaining such eigenvalues
and eigenvectors following the work of \cite{chaos}, rendering our 
semi-analytical approach developed below more proximal to a fully
analytical one.
The overlap integrals 
$A_{0}=\int \Gamma(x)u_{0}^{4}(x)dx$, 
$A_{1}=\int \Gamma(x)u_{1}^{4}(x)dx$, 
$B=\int \Gamma(x)u_{0}^{2}(x)u_{1}^{2}(x)dx$, 
$D_{0}=\int \Gamma(x)u_{0}^{3}(x)u_{1}(x)dx$ and $D_{1}=\int \Gamma(x)u_{0}(x)u_{1}^{3}(x)dx$ are constants. 
Notice that $u_{0}(x)$ and $u_{1}(x)$ are both real functions 
(due to the Hermitian nature of the underlying linear Schr{\"o}dinger problem) and are also orthonormal. In what follows, in the definition of $\Gamma(x)$
in Eq. (\ref{Gamma}), we will set
%
\begin{equation}
\alpha+\beta=-1,\quad \alpha-\beta=-1+\varepsilon,
\label{eq11}
\end{equation}
with $0\le \varepsilon \le2$, as shown in Fig. \ref{fig1}. Notice that in the limiting case 
$\varepsilon=0$, 
the nonlinearity coefficient takes a constant value, $\Gamma(x)=-1$, and the problem is reduced 
to the study of a collisionally homogeneous condensate with 
attractive interatomic interactions in a double well potential; this problem was studied in detail in Ref. \cite{theo} 
by means of the considered two-mode approximation for both symmetric and asymmetric double-well potentials. 
As a result of this choice and of fixing the rest of the parameters,
the above overlap integrals $A_0$, $A_1$, $B$, $D_0$ and $D_1$ are 
functions of $\epsilon$ and their dependence on this
parameter is shown in Fig. \ref{fig1a}. 

We now 
introduce in Eqs. (\ref{eq5})-(\ref{eq6})
the amplitude-phase variables $c_{j}=\rho_{j}e^{i\varphi_{j}}$, $j=0,1$ 
(the functions $\rho_{j}$ and $\varphi_{j}$ are assumed to be real and 
are time-dependent) 
and derive from Eqs. (\ref{eq5}) and (\ref{eq6}) the equations for $\rho_{0}$ and $\varphi_0$:
\begin{eqnarray}
\dot{\rho_{0}}&=& B\rho_{0}\rho_{1}^{2}\sin(2\varphi)
+(D_{0}\rho_{0}^{2}\rho_{1}+D_{1}\rho_{1}^{3})\sin(\varphi),
\label{eq7}
\\
\dot{\varphi_0}&=& (\mu-\omega_{0})-A_{0}\rho_{0}^{2}-B(2\rho_{1}^{2}+\rho_{1}^{2}\cos(2\varphi))
-(3D_{0}\rho_{0}\rho_{1}+D_{1}\frac{\rho_{1}^{3}}{\rho_{0}})\cos(\varphi), 
\label{eq8}
\end{eqnarray}
%
where $\varphi =\varphi_{0}-\varphi_{1}$ 
is the relative phase between the two modes. The equations for $\rho_{1}$ and $\varphi_{1}$ can directly be 
obtained by interchanging indices $1$ and $0$ in Eqs. (\ref{eq7}) and (\ref{eq8}). 

Focusing on real solutions of Eq. (\ref{eq1}), 
we consider the steady solutions, i.e., $\dot{\rho_{0}}= \dot{\varphi_0} =0$ [associated to 
the fixed points of the dynamical system of Eqs. (\ref{eq7})-(\ref{eq8})], with $\varphi=k\pi$ with $k$ an integer. 
In such a case, the equations for $\rho_{0}$ [Eq. (\ref{eq7})] and $\rho_{1}$ 
are automatically satisfied, while the equations for $\varphi_{0}$ [Eq. (\ref{eq8})] 
and $\varphi_{1}$ are reduced to 
the following  algebraic system:
\begin{eqnarray}
(\mu-\omega_{0})-A_{0}\rho_{0}^{2}-3B\rho_{1}^{2}-3D_{0}\rho_{0}\rho_{1}
-D_{1}\frac{\rho_{1}^{3}}{\rho_{0}}&=& 0,
\label{eq9}
\\
(\mu-\omega_{1})-A_{1}\rho_{1}^{2}-3B\rho_{0}^{2}-3D_{1}\rho_{1}\rho_{0}
-D_{0}\frac{\rho_{0}^{3}}{\rho_{1}}&=& 0.
\label{eq10}
\end{eqnarray}
%
The semi-analytical part of our considerations will consist of solving these
algebraic conditions for given linear potential $V(x)$ and nonlinear
spatial dependence $\Gamma(x)$ parameters (as mentioned above) and we will compare the 
findings from this two-mode, Galerkin-type truncation with the full
numerical results in what follows.

It is interesting to briefly address the case where $\phi\neq k \pi$
(hence $n(\phi) \neq 0$). In that case, the stationary form of
Eq. (\ref{eq7}) leads to: 
\begin{eqnarray}
\cos(\phi) = - \frac{D_0 \rho_0^2 + D_1 \rho_1^2}{2 B \rho_0 \rho_1}.
\label{add1}
\end{eqnarray}
Then it is straightforward to observe that since $D_0 < D_1$, for all
$\epsilon$, and $B < D_1$ for $\epsilon > 0.995$ (from Fig. \ref{fig1a}), 
for such values
of $\epsilon$, the fraction of the right-hand-side of Eq. (\ref{add1})
can be shown to be $<-1$ and, hence, there are no other solutions
satisfying Eq. (\ref{eq7}). For $\epsilon < 0.995$ we could not perform 
a similar
 semi-analytical calculation. However, we confirmed numericallly 
the non-existence of 
additional solutions (to the ones with $\phi= k \pi$) for this domain.

\section{Numerical Results}

\begin{figure}[tbhp!]
     \centering
           \includegraphics[width=.4\textwidth]{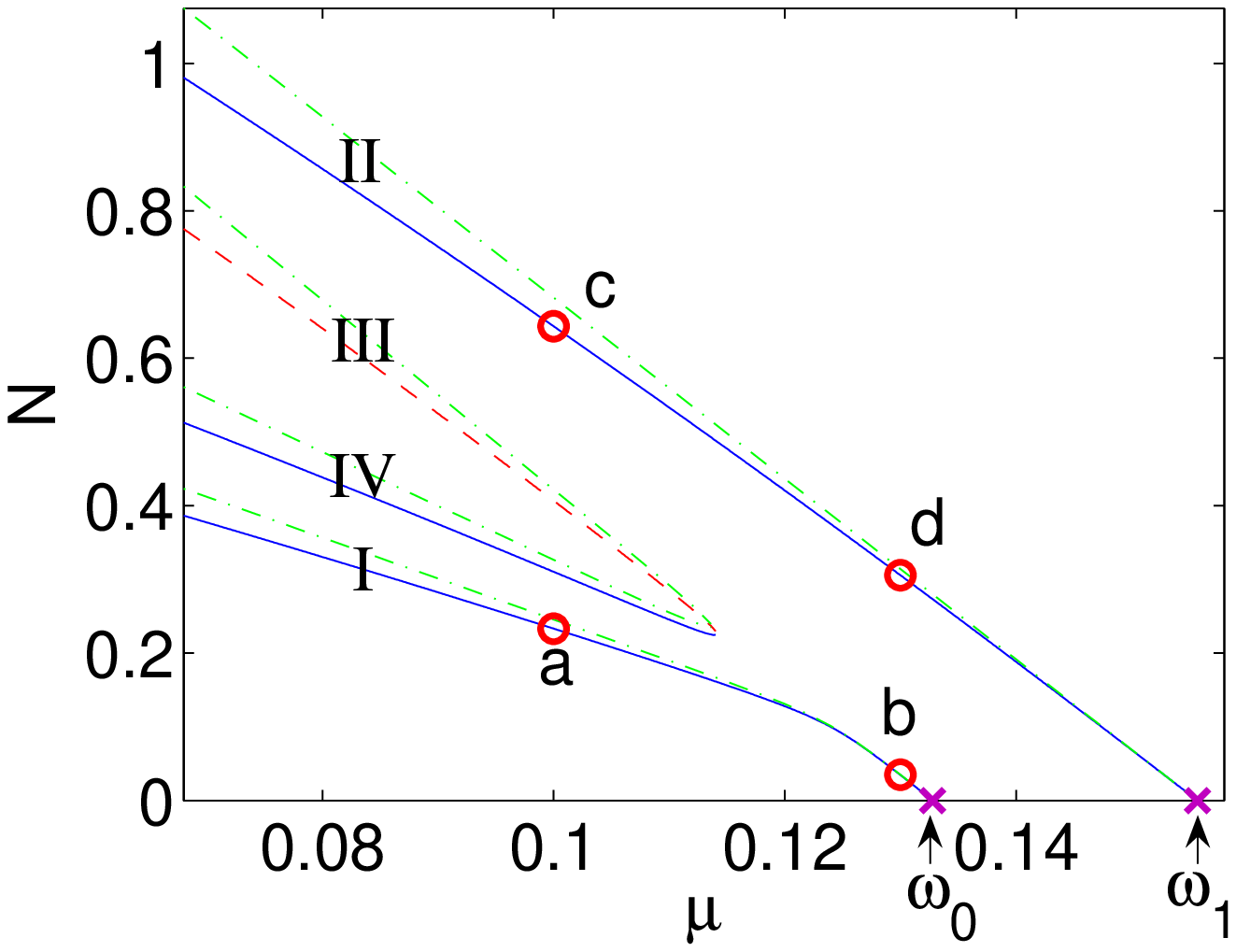}
           \includegraphics[width=.4\textwidth]{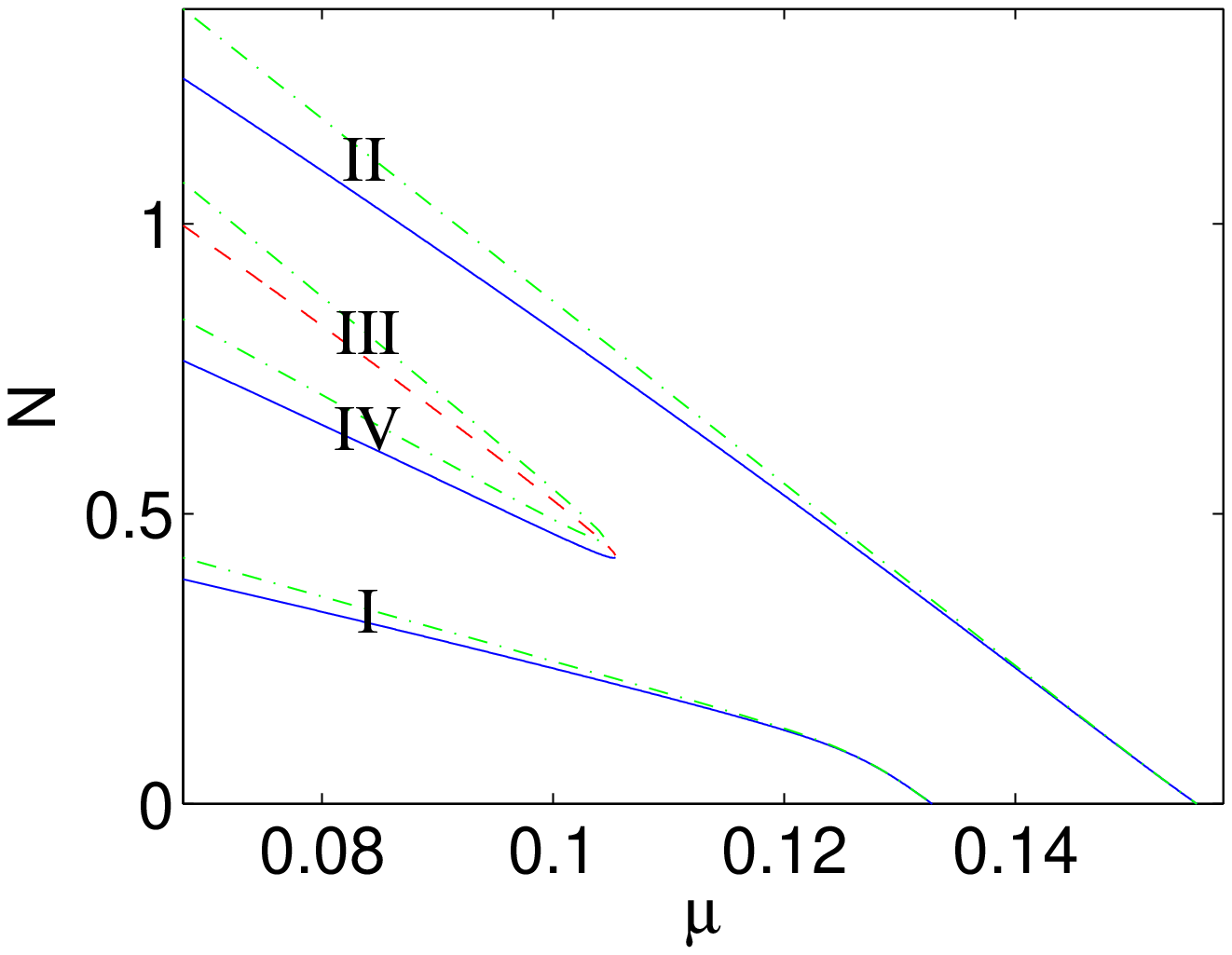}\\
           \includegraphics[width=.2\textwidth]{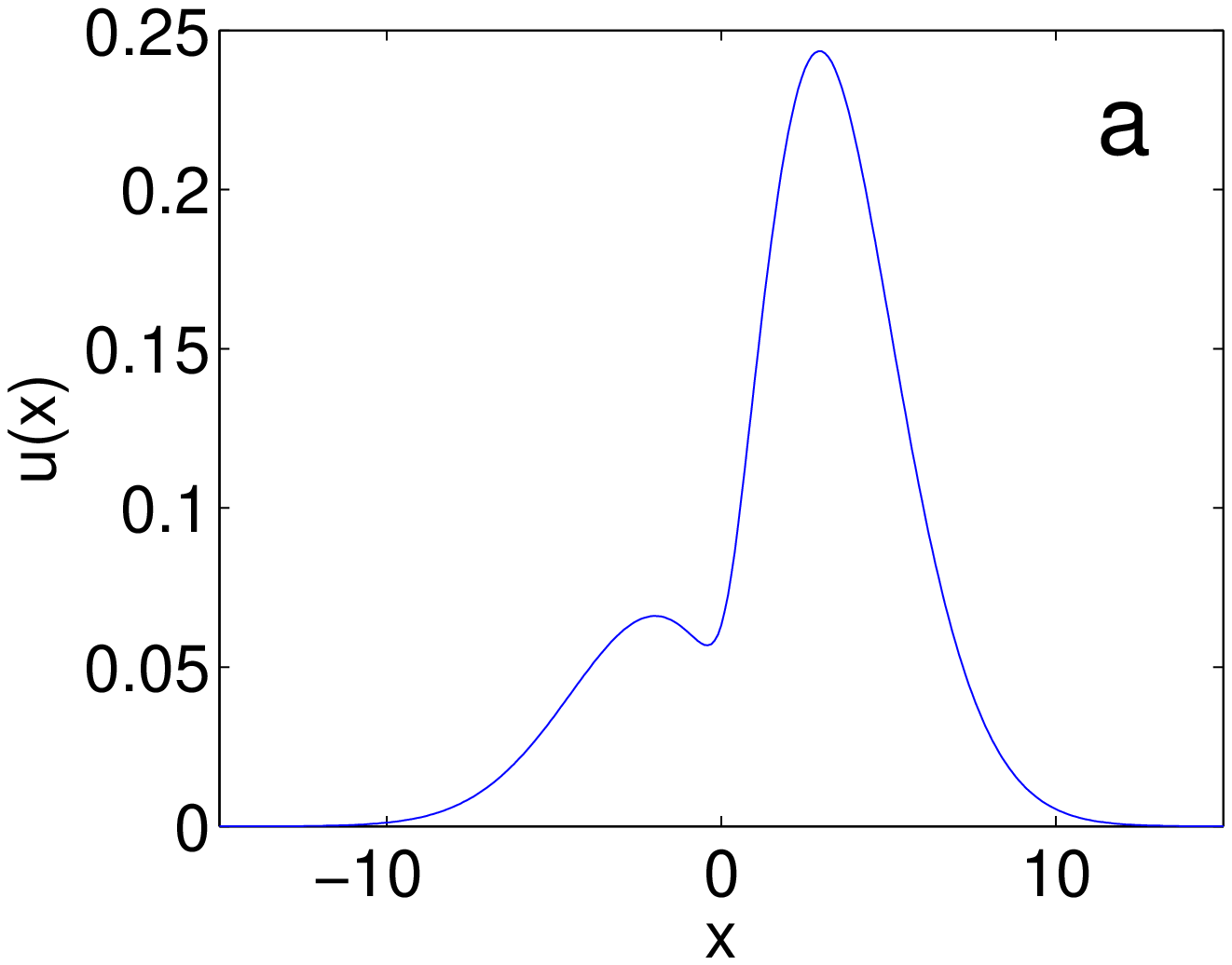}
           \includegraphics[width=.2\textwidth]{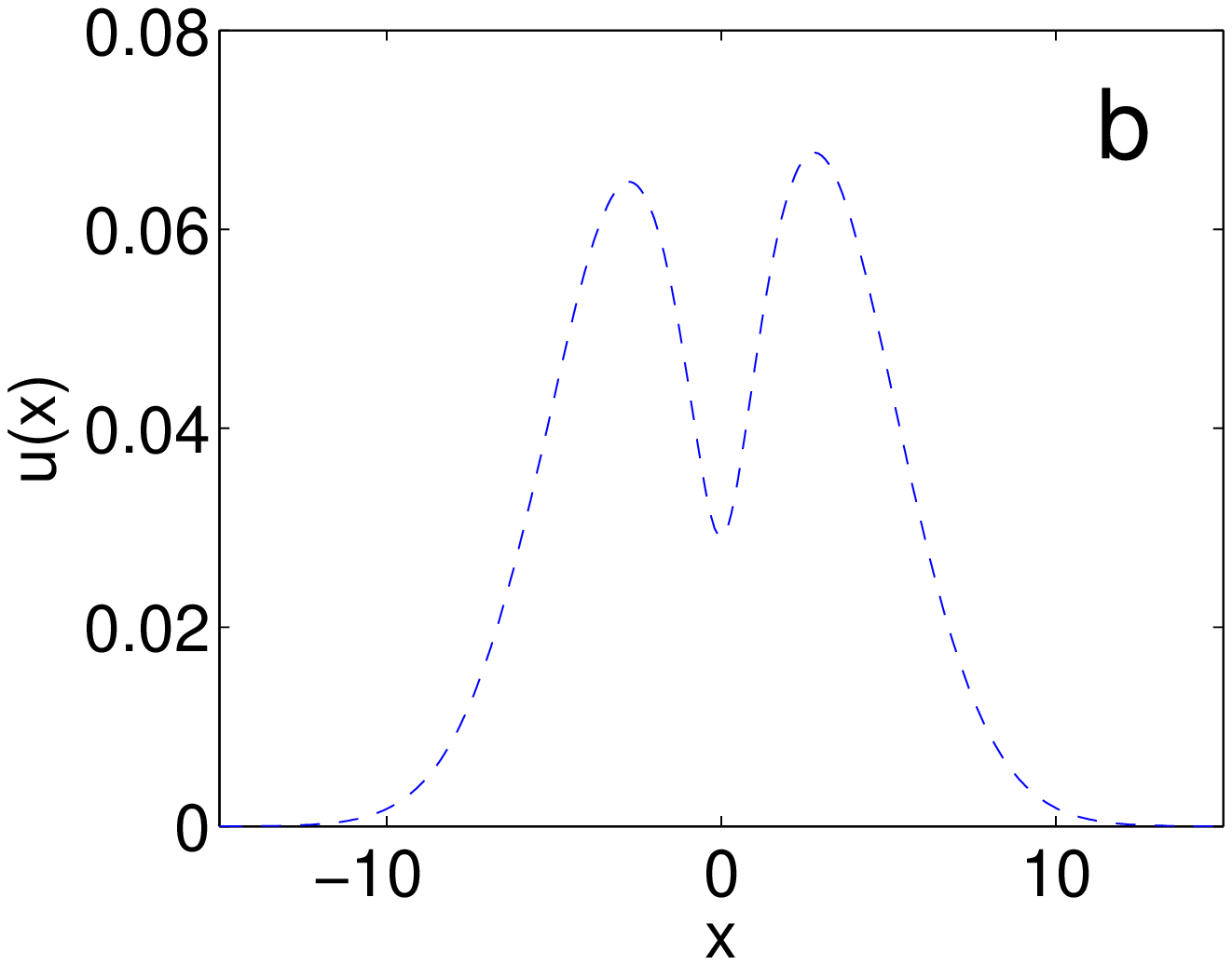}\\
           \includegraphics[width=.2\textwidth]{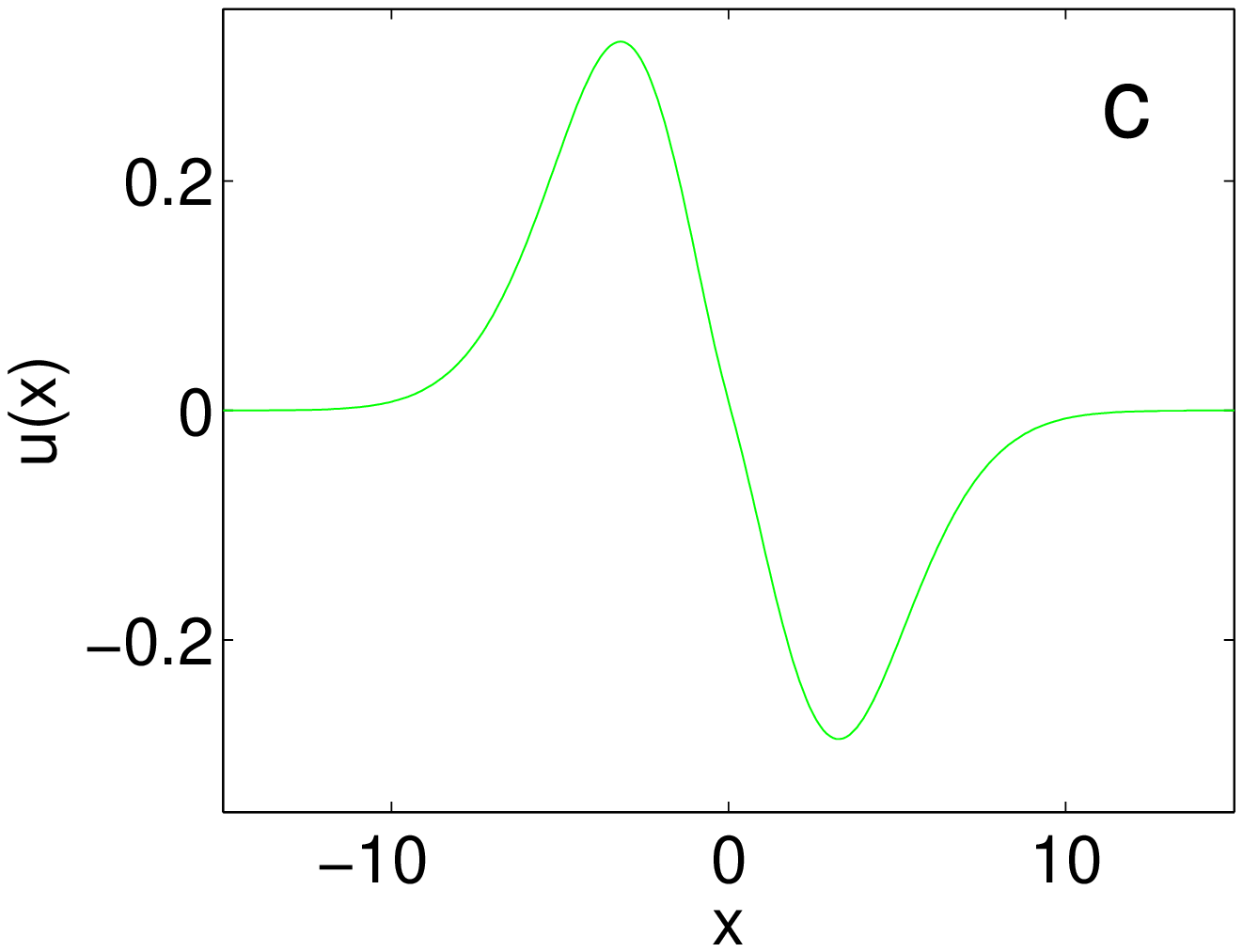}
           \includegraphics[width=.2\textwidth]{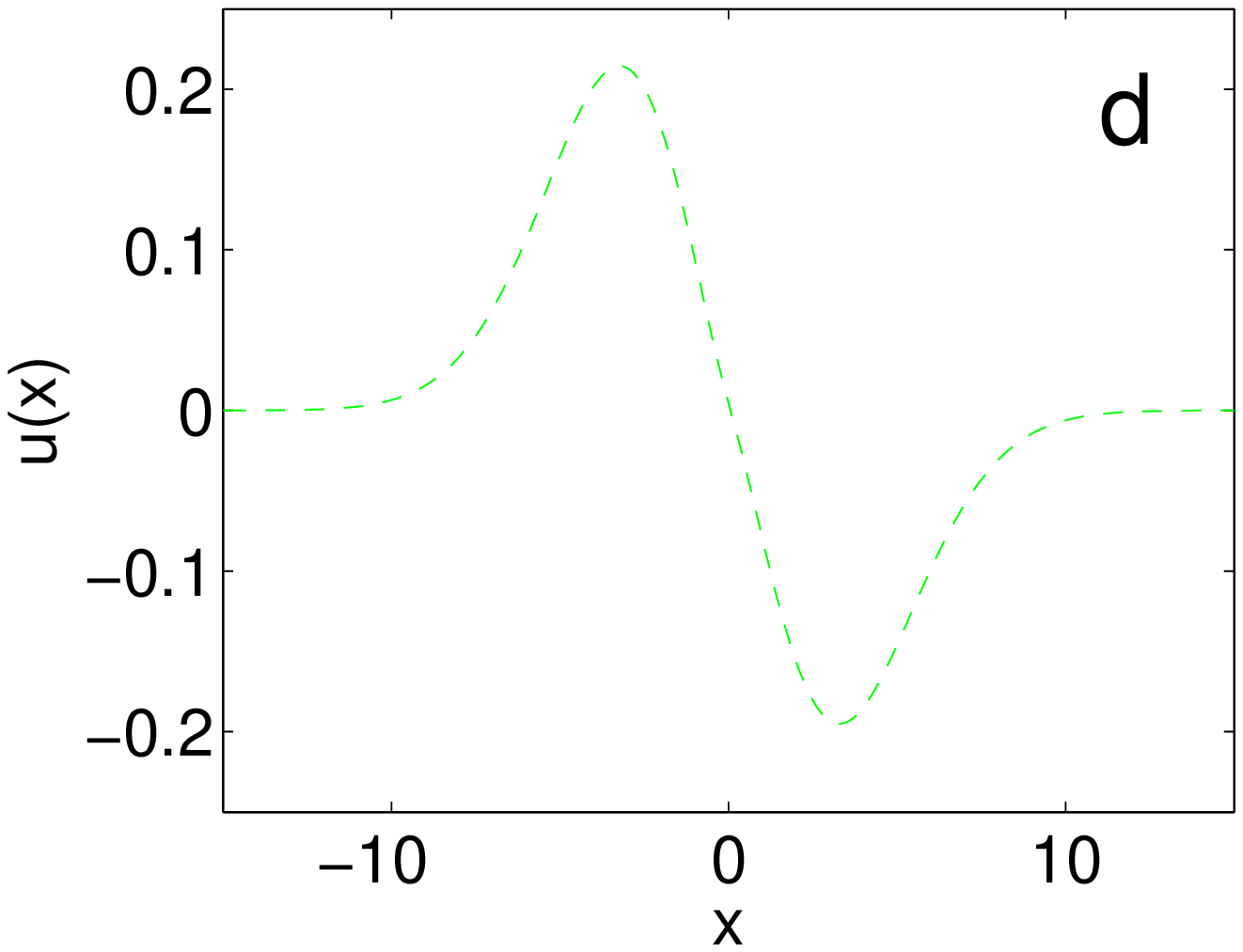}\\
      \caption{Top panels: The normalized number of atoms $N$ [see Eq. (\ref{N})] 
of the solutions of Eq. (\protect\ref{eq1}) for the case of attractive interatomic interactions 
i.e., for a nonlinearity coefficient $\Gamma(x)$ 
with parameters $\alpha+\beta=-1$ and $\alpha-\beta=-1+\varepsilon$, with $\varepsilon=0.25$ (top left) 
and $\varepsilon=0.5$ (top right), as a function of the normalized chemical potential $\mu$. 
The blue solid lines and red dashed lines denote the stable and unstable numerically found solutions. 
The green dashed-dotted lines depict the result of the two-mode approximation. 
Notice that as $N \rightarrow 0$, the spatial profiles of the two branches
tend to the linear eigenmodes $u_{0,1}$ and accordingly $\mu \rightarrow
\omega_{0,1}$.
Bottom panels: 
The profiles of the wave functions corresponding to branch I (upper blue) and II (lower green). 
Along each branch, the profiles are 
shown for two values of $\mu$, i.e., $\mu=0.1$ (left solid) and $\mu=0.13$ (right dashed); the corresponding labels are a) and b) for the symmetric (in the
linear limit) branch shown above, while c) and d) are for the antisymmetric
(in the linear limit) branch shown below.}
      \label{fig2}
\end{figure}

We begin the exposition of our numerical 
results by considering $0<\varepsilon<1$, in which case $\Gamma(x)<0$, i.e., 
attractive interatomic interactions. The 
top panels of Fig. \ref{fig2} show two examples: $\varepsilon=0.25$ (left) and $\varepsilon=0.5$ (right). 
Each of these panels 
presents the complete diagram of the numerically generated solutions (blue solid lines 
correspond to stable solutions and red dashed lines to unstable 
ones) to 
Eq. (\ref{eq1}) and the corresponding analytical predictions (green dashed-dotted lines), namely, 
the results of the two mode approximation obtained by solving Eqs. (\ref{eq9}) and (\ref{eq10}). 
The solutions are expressed in terms of the normalized number of atoms $N$ [see Eq. (\ref{N})] 
as a function of the normalized chemical potential $\mu$. The branches are obtained 
using a fixed-point Newton iteration and numerical continuation from the non-interacting 
limit of the system (i.e., in the absence of nonlinearity).
Then, numerical linear stability analysis is performed to determine whether each branch
does or does not possess real eigenvalues (which result in instability). 
It is readily observed that the analytical predictions for the
different branches through Eqs. (\ref{eq9})-(\ref{eq10}) are in very 
good agreement with the numerically found solutions.

\begin{figure}[tbhp!]
     \centering
           \includegraphics[width=.4\textwidth]{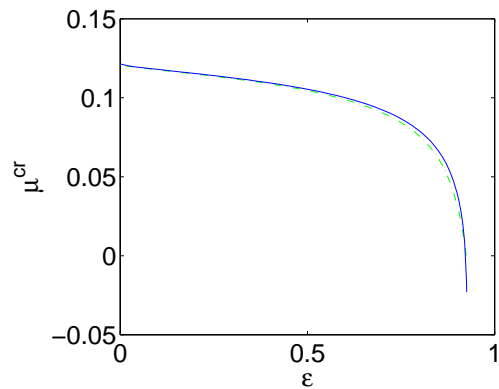}
      \caption{The value of $\mu$ at which branches III and IV disappear as a function of 
$\varepsilon$ with respect to Fig. \protect\ref{fig2}. The blue solid lines and the green dashed-dotted lines 
denote the numerically found solutions and the result of the two-mode approximation, respectively.}
      \label{fig3}
\end{figure}

Let us now discuss the various branches appearing in the bifurcation diagrams in more detail. 
Branch I corresponds to the asymmetric (due to the collisional inhomogeneity)
solution starting at $\mu=\omega_{0}$, the eigenvalue corresponding to the 
symmetric ground eigenstate $u_{0}$. 
The asymmetry arises immediately after the deviation from the linear
limit of $N\to 0$, and becomes increasing as one drifts further away.
Similarly, branch II starts from $\mu=\omega_{1}$, the eigenvalue of the 
anti-symmetric first excited eigenstate, and 
becomes increasingly asymmetric as it gets away from its linear limit, i.e., 
when $N$ gets larger. The 
four bottom panels of Fig. \ref{fig2} 
provide specific examples of the profiles of this continuation.

Branches III and IV correspond to a pair of two other asymmetric solutions 
which collide at some critical value of $\mu \equiv \mu^{cr}$  
and disappear from then on through a saddle-mode bifurcation. As for the two  
examples in Fig. \ref{fig2}, the critical points are 0.1140 and 0.1054, for the 
cases $\varepsilon=0.25$ and $\varepsilon=0.5$, respectively. Moreover, branch IV is observed to move 
towards branch I as $\varepsilon \to 0$.
Specifically, when $\varepsilon=0$, branch IV 
merges into branch I and the diagram turns into a 
pitchfork bifurcation (the two asymmetric branches are mirror images 
of each other and both bifurcate from the symmetric solution in that
case), which is the 
case with attractive interactions 
analyzed in Ref. \cite{theo}. 
As $\varepsilon$ increases, the critical point $\mu^{cr}$ of the
saddle-node bifurcation decreases and tends to 
negative infinity rapidly, especially as $\varepsilon$ gets closer to 1; 
then, branches III and IV keep moving to the left of the 
diagram and disappear finally when $\varepsilon$ is close enough to 1. Also, 
the stability analysis indicates that branch III is the only unstable one among 
the four solutions, while the other three are all stable for any value of $\mu$.

As a stringent test of our analytical two-mode approximation, in 
Fig. \ref{fig3} we 
show the critical point of the saddle-node 
bifurcation $\mu^{cr}$ as a function of the collisional inhomogeneity 
parameter $\varepsilon$. We observe that the solid line of the 
fully numerical results 
is almost identical to the dashed line yielding 
the theoretical prediction for the occurrence of this bifurcation. 
This figure demonstrates excellent agreement between the analytical results and the numerical findings.

\begin{figure}[tbhp!]
     \centering
           \includegraphics[width=.4\textwidth]{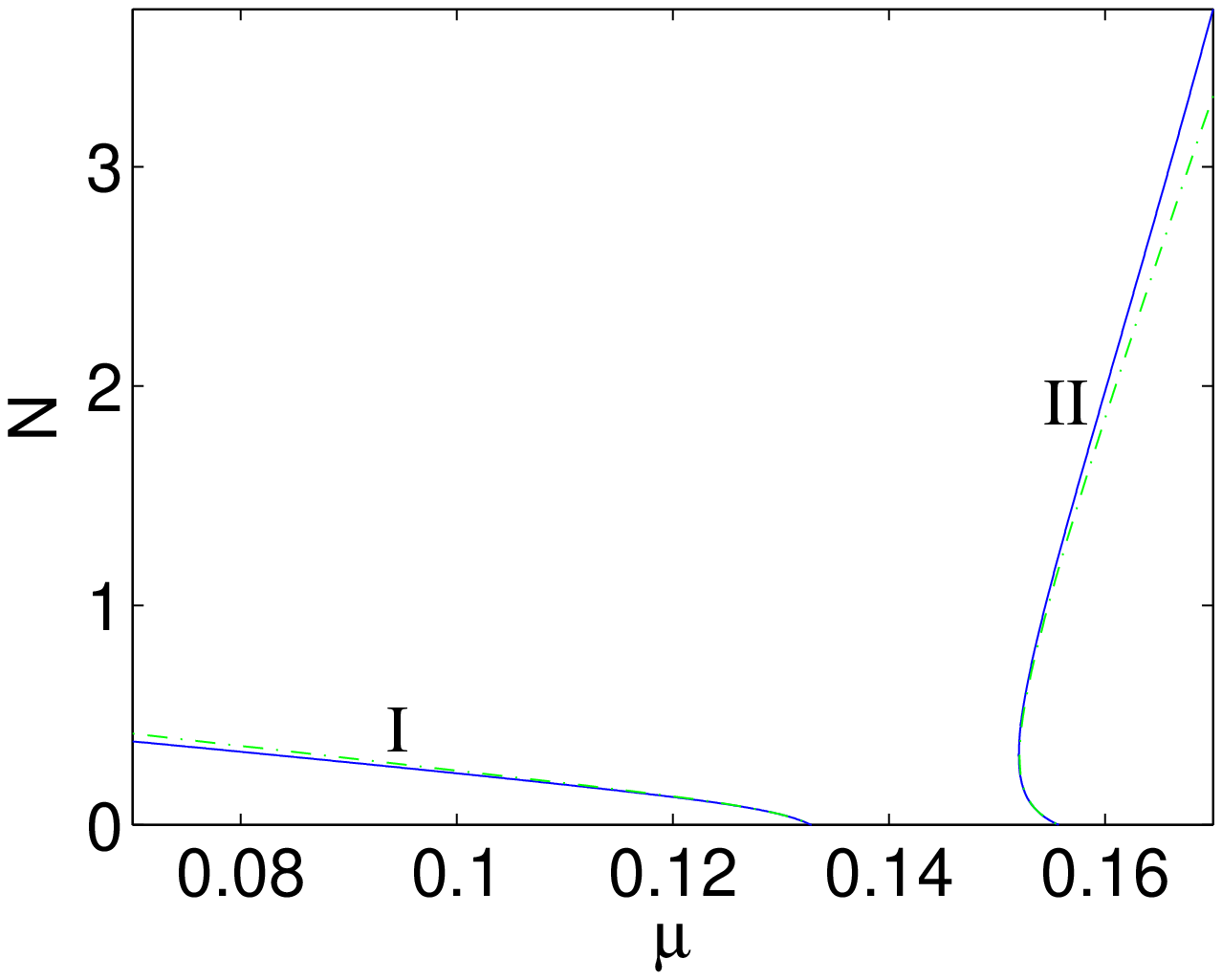}
           \includegraphics[width=.4\textwidth]{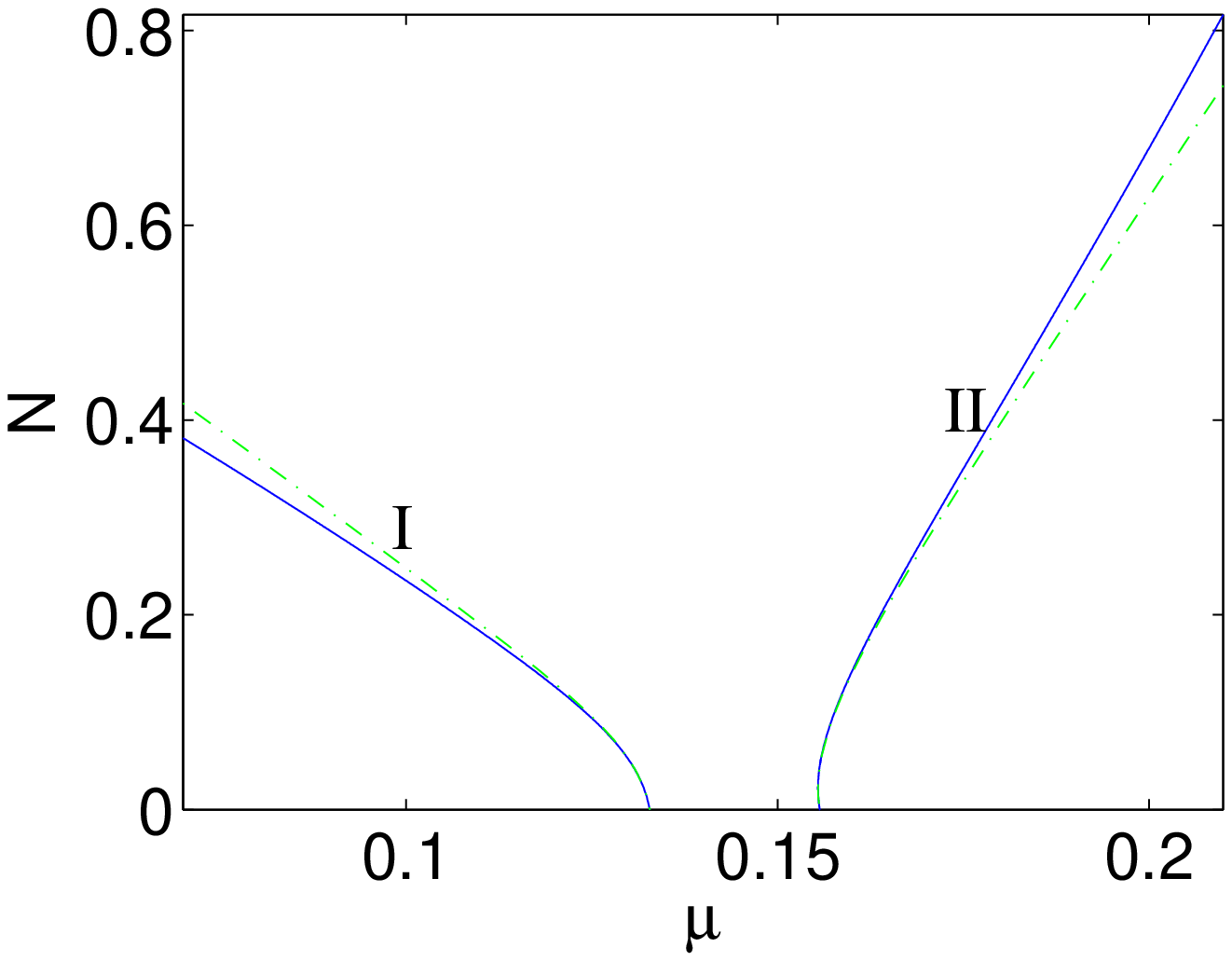}\\
           \includegraphics[width=.4\textwidth]{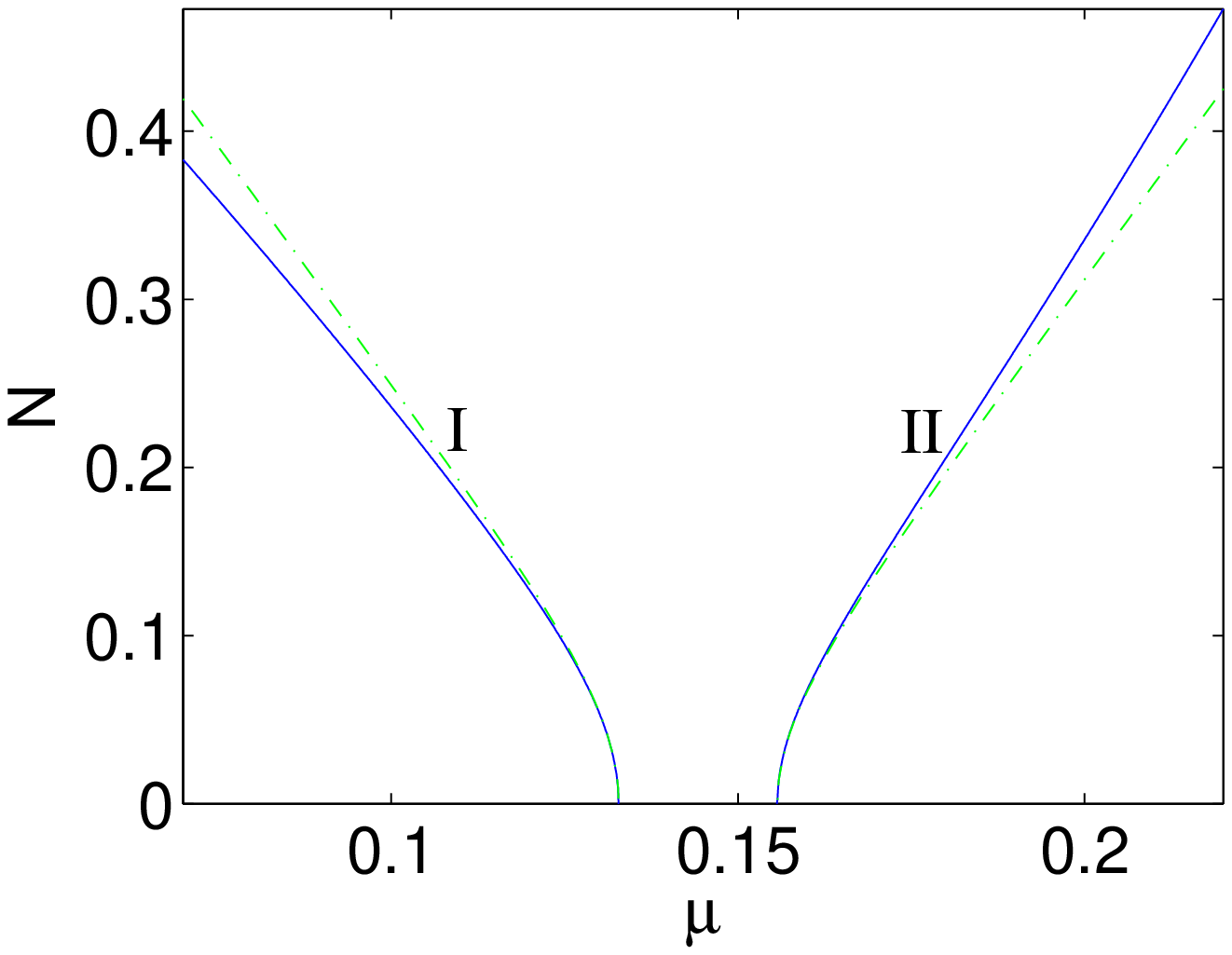}
      \caption{The norm of the solutions of Eq. (\protect\ref{eq1}) for 
a nonlinearity coefficient $\Gamma(x)$ with parameters such that 
$\alpha+\beta=-1$ and $\alpha-\beta=-1+\varepsilon$, with $\varepsilon=1.05$ (top left), 
$\varepsilon=1.5$ (top right) and $\varepsilon=2$ (bottom), as a function of $\mu$. 
The notation is the same as in Fig. \protect\ref{fig2}.}
      \label{fig4}
\end{figure}

Next, we 
study 
the behavior of the solutions when $\varepsilon > 1$, 
in which case the nonlinearity is no longer purely attractive. 
Remarkably, in this case, only two solutions, branch I and II, still survive 
for all the examined values of $\varepsilon > 1$.
We realize that when $\varepsilon$ grows larger from 0 to 1, both branches I and II 
move 'clockwise' and branch II appears to be almost vertical when $\varepsilon$ 
passes the value 1. 
As $\varepsilon$ continues to increase from 1 to 2, we find an 
interesting new phenomenon: 
within a certain small range of $N$, the 
solution II only exists for chemical potentials $\mu$ slightly less than the eigenvalue $\omega_{1}$ 
(corresponding to the eigenstate $u_1$) 
before it meets a turning point; the latter occurs, for example, at $\mu=0.1520$ 
or $0.1554$ for the cases $\varepsilon=1.05$ 
or $1.5$, respectively. 
After the turning point, the solution exists when $\mu$ gets larger. This phenomenon is shown 
in Fig. \ref{fig4}, in which 
it is observed that branch II 
starts at the linear limit, 
persists with $dN/d \mu < 0$ for a narrow interval of chemical potentials,
before turning to the right with and acquiring $dN/d\mu>0$. 
It is also worth mentioning that both branches I and II are still 
asymmetric and {\it stable} in this case; 
thus, here we observe an interesting deviation from the 
well-known Vakhitov-Kolokolov criterion \cite{vk}
(see, e.g., 
a relevant discussion in Ref. \cite{boris}) about the slope of the branch 
determining its linear stability. We present an explanation of
the relevant feature at the end of this section
 (after observing similar features in the principally defocusing case below).

\begin{figure}[tbhp!]
     \centering
           \includegraphics[width=.4\textwidth]{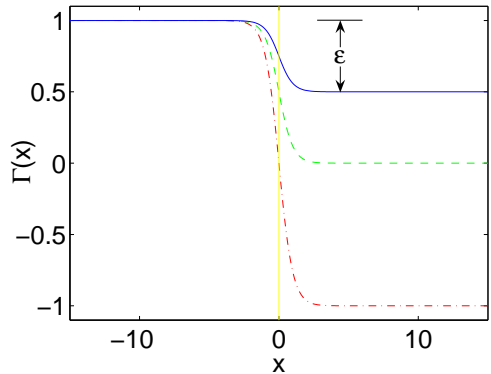}
     \caption{Similar to Fig. \ref{fig1}, but with a nonlinearity coefficient $\Gamma(x)$ such that 
$\alpha+\beta=1-\varepsilon$, $\alpha-\beta=1$ and $b=1$, with $\varepsilon=0.5$ (blue solid line), 
$\varepsilon=1$ (green dashed line) and $\varepsilon=2$ (red dashed-dotted line).}
      \label{fig5}
\end{figure}

Now let us consider the case in which the parameters involved in the nonlinearity coefficient $\Gamma(x)$ are such that 
%
\begin{equation}
\alpha+\beta=1-\varepsilon,\quad \alpha-\beta=1,
\label{eq12}
\end{equation}
with $0\le \varepsilon \le2$, as shown in Fig. \ref{fig5}. 
Notice that Eqs. (\ref{eq11}) and (\ref{eq12}) produce the same results when $\varepsilon=2$. 

\begin{figure}[tbhp!]
     \centering
           \includegraphics[width=.4\textwidth]{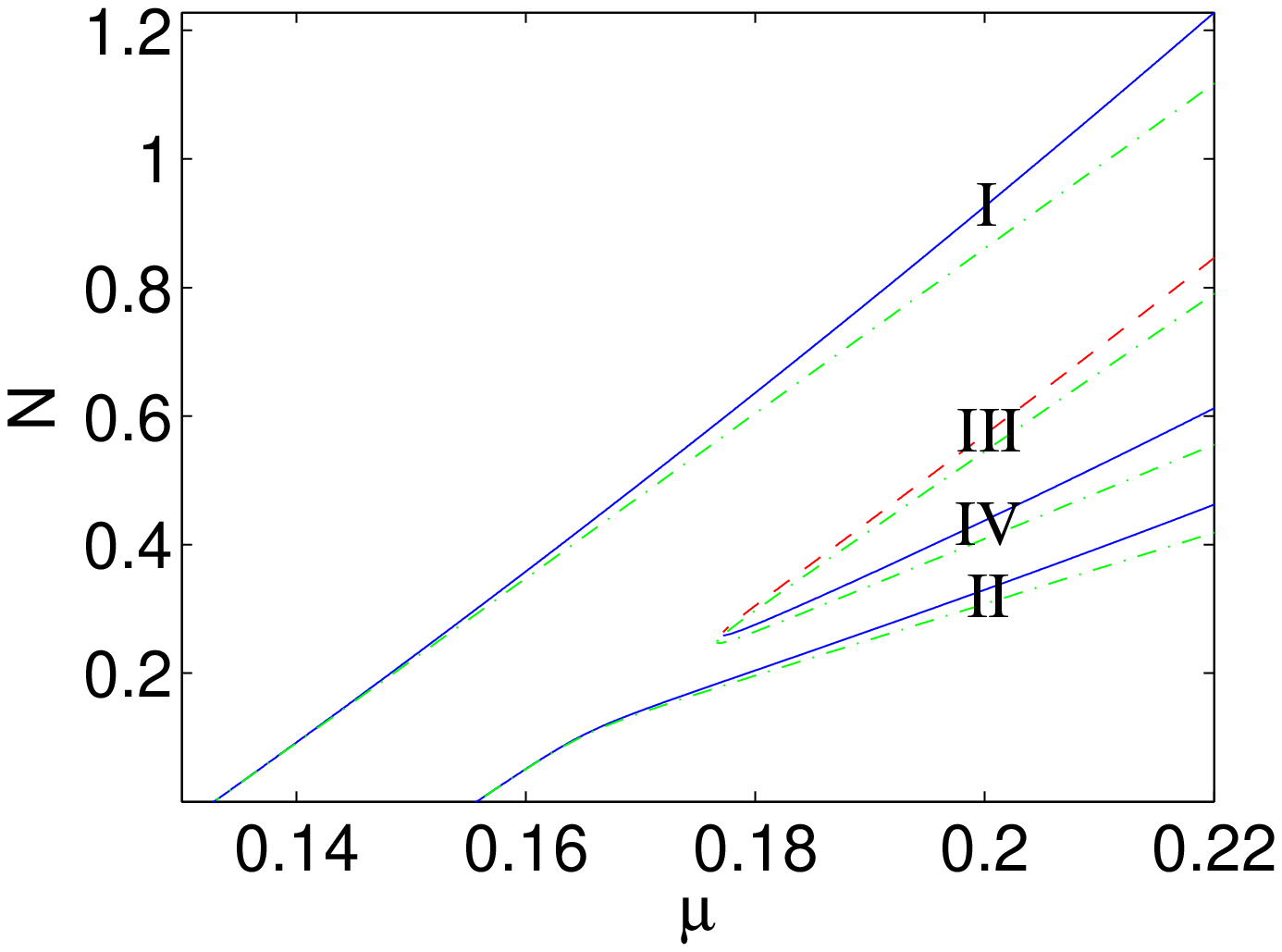}
           \includegraphics[width=.4\textwidth]{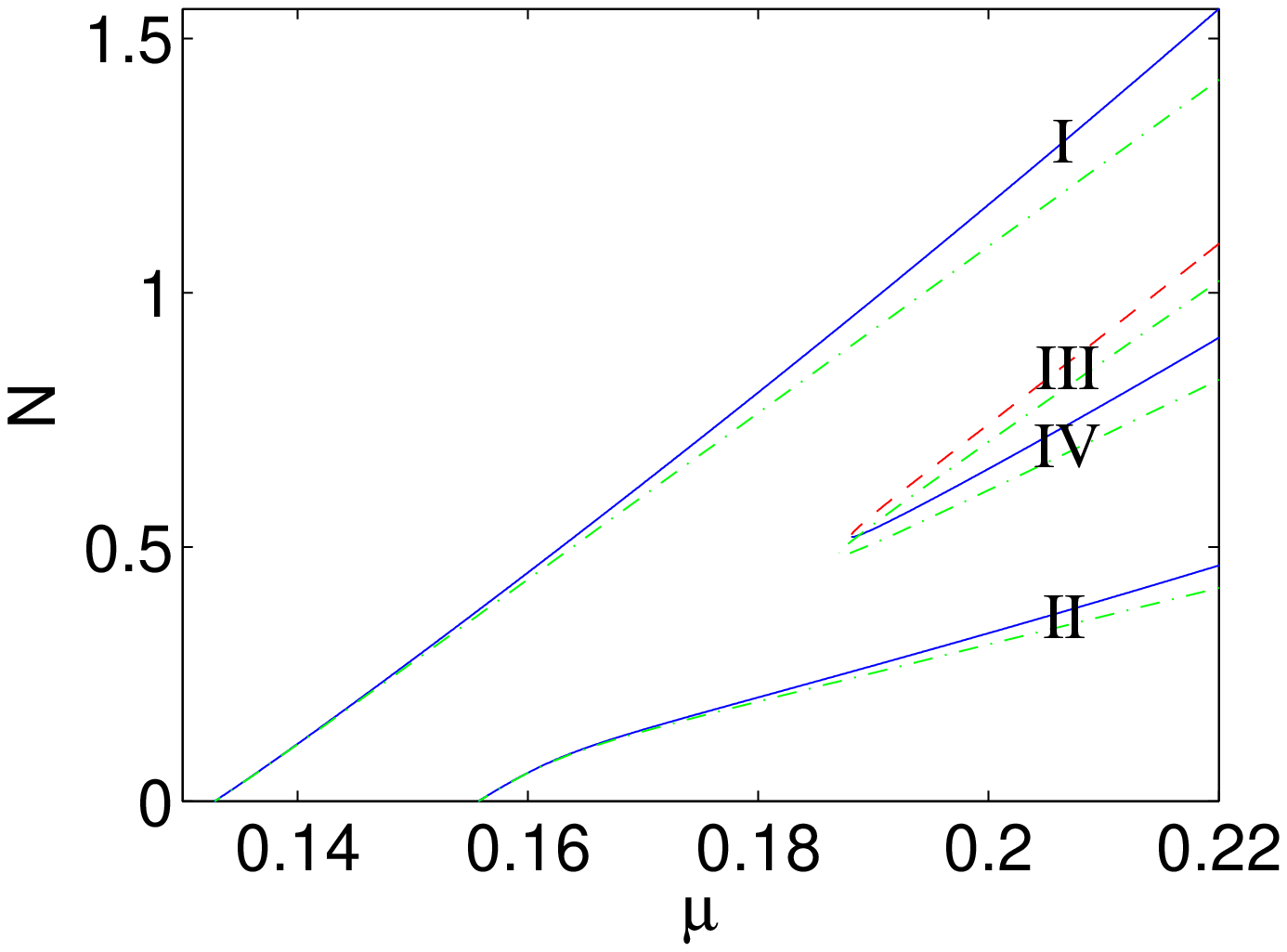}\\
      \caption{The normalized number of particles $N$ for the solutions 
of Eq. (\protect\ref{eq1}) as a function of $\mu$ 
in the case of repulsive interatomic interactions, 
namely, for a nonlinearity coefficient $\Gamma(x)$ 
with parameters such that $\alpha+\beta=1-\varepsilon$ and $\alpha-\beta=1$, 
with $\varepsilon=0.25$ (left) and $\varepsilon=0.5$ (right).
The notation is the same as in Fig. \protect\ref{fig2}.}
      \label{fig6}
\end{figure}

\begin{figure}[tbhp!]
     \centering
           \includegraphics[width=.4\textwidth]{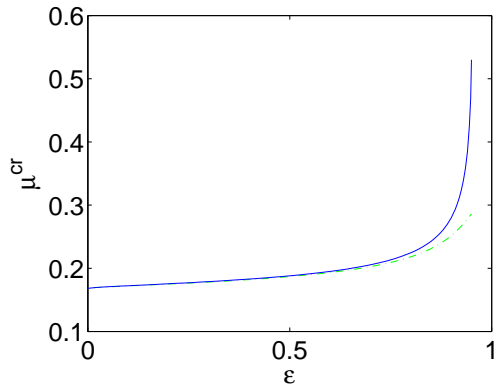}
      \caption{The critical value $\mu^{cr}$ of 
the normalized chemical potential at which branches III and IV disappear as a function of $\varepsilon$ 
with respect to Fig. \protect\ref{fig5}. The blue solid lines and the green dashed-dotted lines denote the 
numerically found solutions and the prediction of the two-mode approximation, respectively.}
      \label{fig7}
\end{figure}

Figure \ref{fig6} shows some prototypical examples of the bifurcation diagram 
of the relevant solutions when $0<\varepsilon<1$, 
i.e., in the case of a purely repulsive nonlinearity (notice that 
the branches turn to the opposite direction than
in Fig. \ref{fig2}). Similar to the previous case, four solutions are found, 
two of which disappear through a saddle node bifurcation, allowing only two
to survive in the non-interacting (linear) limit of $N \rightarrow 0$.
As before, the saddle-node bifurcation becomes a pitchfork one 
(but now emerging from the anti-symmetric branch) in the case
of a collisionally homogeneous environment i.e., for $\varepsilon \rightarrow 0$. 
Preserving the original notations, branches I and II are the two asymmetric solutions 
extending to the linear limit, and 
starting at $\mu=\omega_{0}$ and $\mu= \omega_{1}$, respectively. 
Once again branches III and IV disappear when $\mu$ decreases to some 
critical value $\mu^{cr}$, which goes up to infinity as $\varepsilon$ increases to 1. 
The dependence of $\mu^{cr}$ on the inhomogeneity parameter $\varepsilon$ is shown 
in Fig. \ref{fig7}. It is important once again to highlight the 
good qualitative and even quantitative 
(apart from the case of very large $\varepsilon$) 
agreement of the two-mode prediction for $\mu^{cr}$ with the full
numerical results.

%

\begin{figure}[tbhp!]
     \centering
           \includegraphics[width=.4\textwidth]{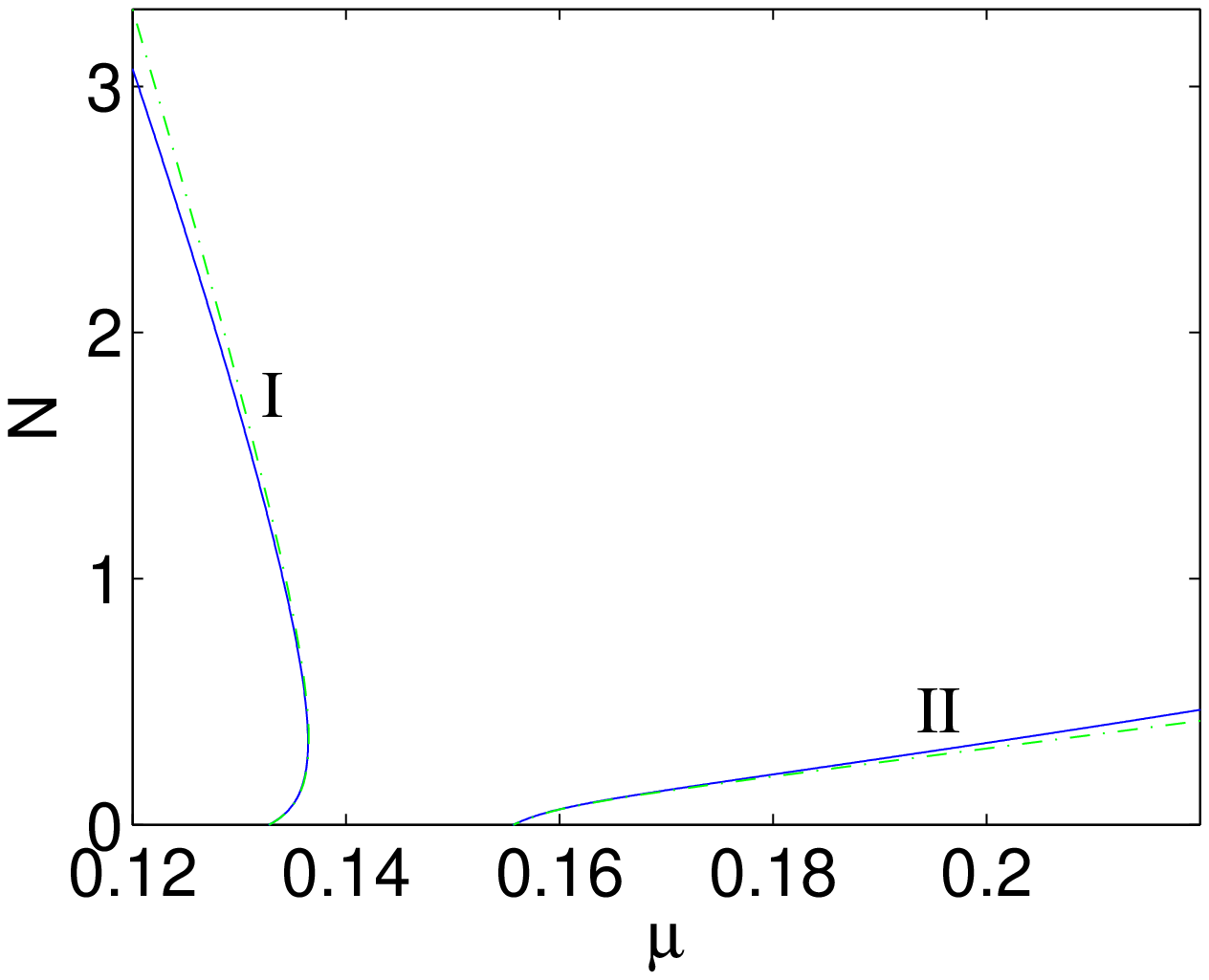}
           \includegraphics[width=.4\textwidth]{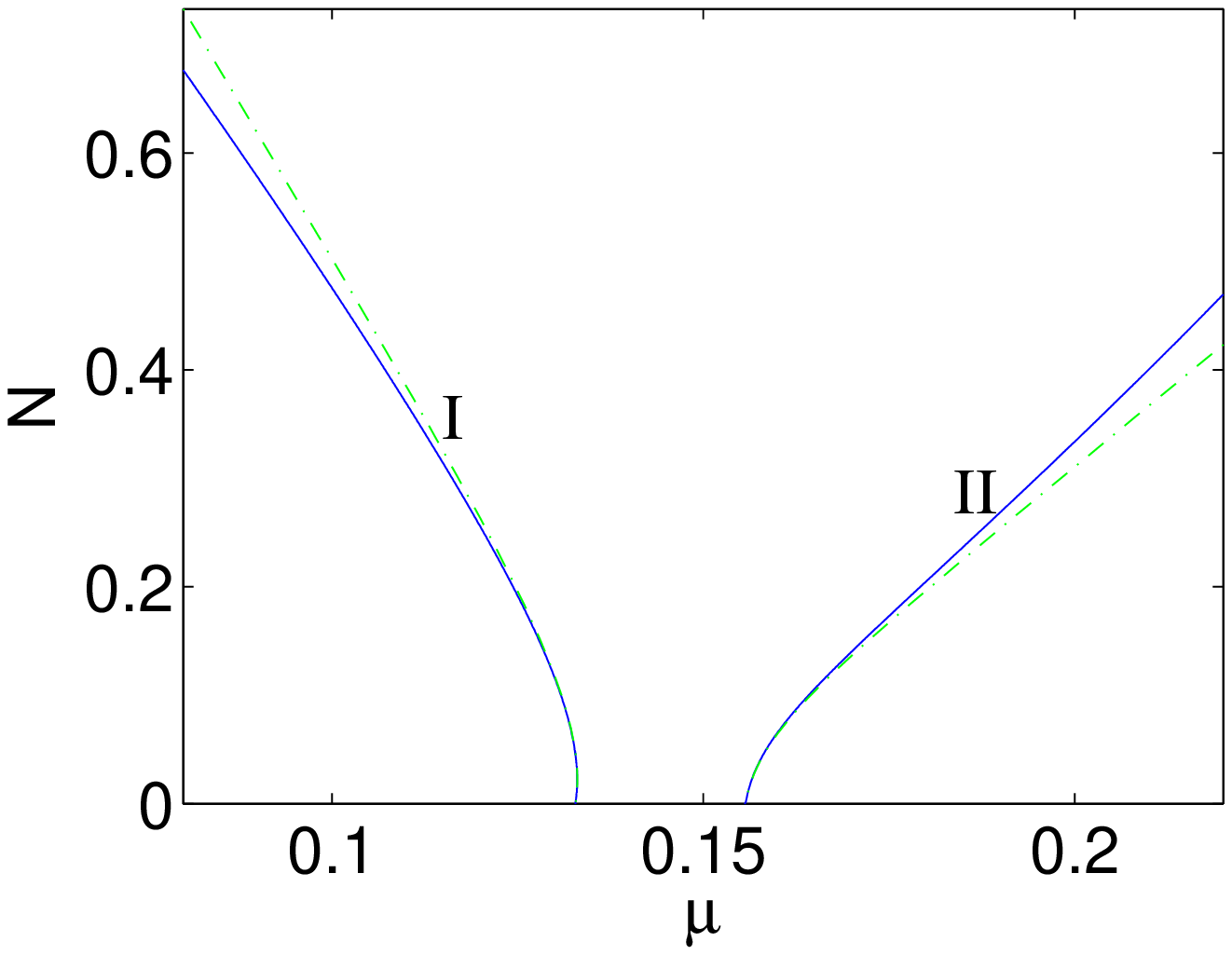}\\           
      \caption{The 
normalized number of particles $N$ of the solutions of Eq. (\protect\ref{eq1}) as a function of $\mu$, 
for a nonlinearity coefficient $\Gamma(x)$ with parameters such that 
$\alpha+\beta=1-\varepsilon$ and $\alpha-\beta=1$, with $\varepsilon=1.05$ (left) and $\varepsilon=1.5$ (right), 
The notation is the same as in Fig. \protect\ref{fig2}.}
      \label{fig8}
\end{figure}

It is interesting to note that in this case as $\varepsilon$ increases and 
one of the wells becomes less repulsive (and eventually attractive for $\varepsilon > 1$),  
the branches I and II keep rotating 'counterclockwise'. 
For $\varepsilon > 1$, the turning of the branch arises again, on branch I 
this time, as shown in Fig. \ref{fig8}. In this case, the bifurcation diagram 
contains only two branches which, for sufficiently large $\varepsilon$,  
feature opposite monotonicity of the dependence of $N$ on the chemical
potential $\mu$ (although both branches are linearly stable).
Similarly to the previous case, branch III is the unstable solution, 
while the other three remain stable. 
In all the cases (and even these of large $\varepsilon$),  
we again note 
the strong agreement between 
the bifurcation diagram predicted by the two-mode approximation, in comparison with the 
numerical results.

\begin{figure}[tbhp!]
      \centering
           \includegraphics[width=.4\textwidth]{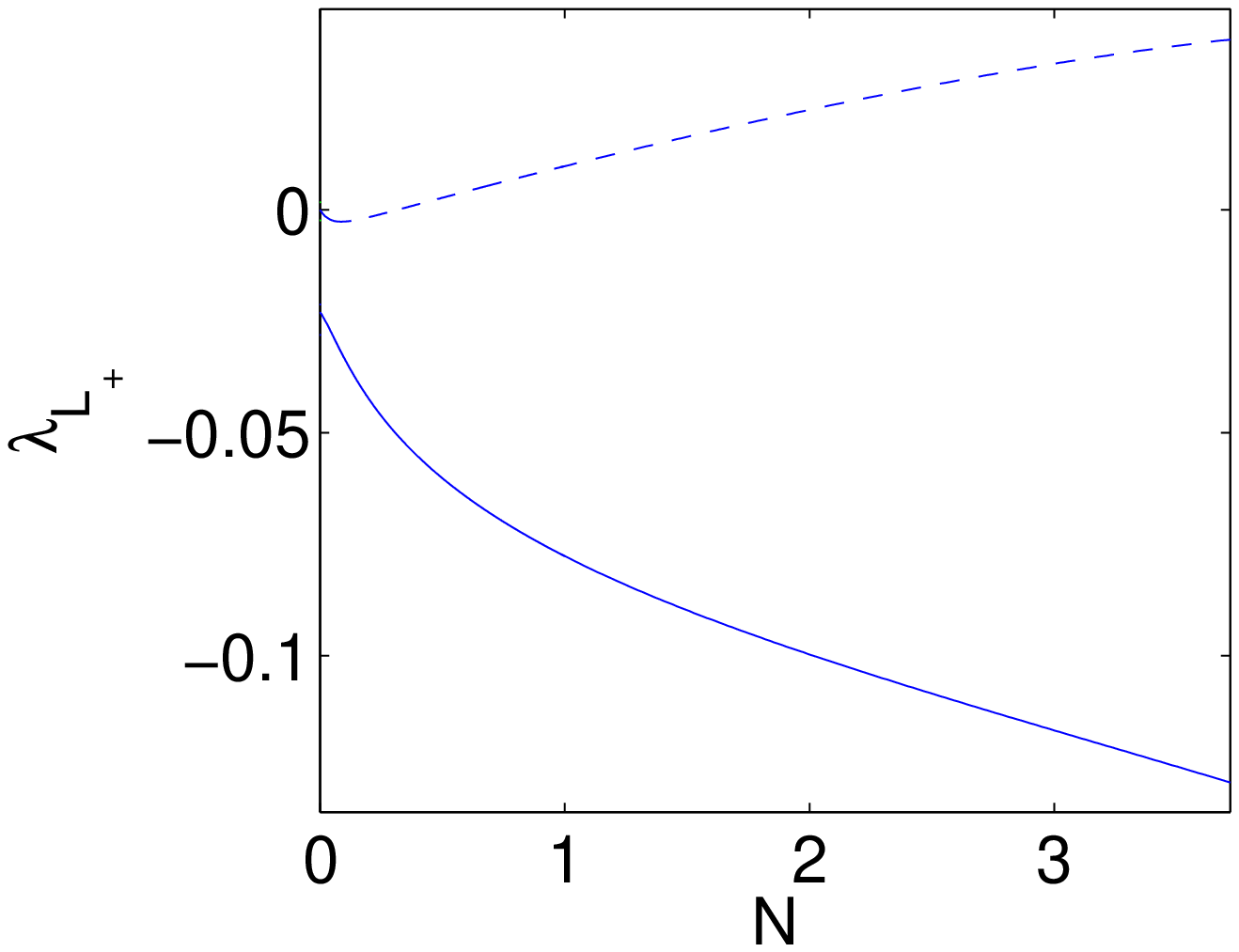}
           \includegraphics[width=.4\textwidth]{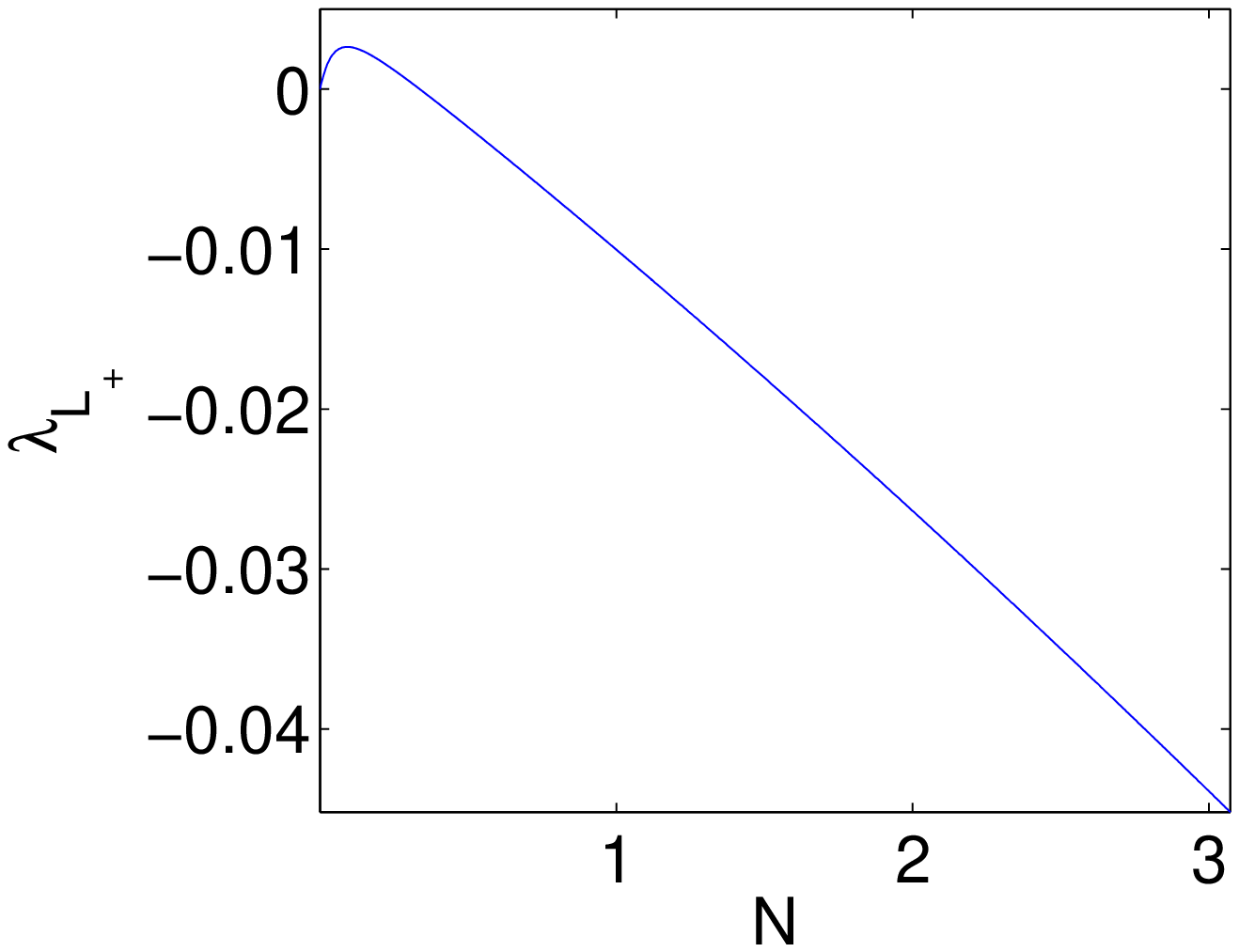}
      \caption{The eigenvalues of the operator $L_{+} = -\frac{1}{2}\partial_{x}^{2} + V(x)
      + 3g|u|^{2} -\mu$, which involve negative parts, as a function of N. The left and right panels relate to the two ``turning branches'', i.e. branch II in Fig. \protect\ref{fig4} and branch I in Fig. \protect\ref{fig8}, respectively, for the case that $\varepsilon=1.05$.}
      \label{figLn}
\end{figure}

We observe that in Fig. \ref{fig4} (e.g., in its top left panel)
for the case where the nonlinearity is principally focusing,
as well as in Fig. \ref{fig8} (e.g., in its left panel),
where it is principally defocusing, one of the branches changes
its monotonicity, as a result of the spatially dependent nonlinearity.
As indicated previously, given the slope condition of the 
Vakhitov-Kolokolov criterion \cite{vk}, it appears to be
rather surprising that this change of monotonicity is not
accompanied by a change of stability. However, we argue here that it is not.
Defining the well known linearization operators
\begin{eqnarray}
L_{+} &=& -\frac{1}{2}\partial_{x}^{2} + V(x)
      + 3 g(x) |u|^{2} -\mu
\label{stuff}
\\
L_{-} &=& -\frac{1}{2}\partial_{x}^{2} + V(x)
      + g(x) |u|^{2} -\mu,
\label{stuff2}
\end{eqnarray}
it is known from the work of \cite{grillakis} that
when $|n(L_+)-n(L_-)|=1$ [i.e., the number of negative eigenvalues
of $L_+$ minus the negative eigenvalues of $L_-$ is in absolute
value equal to 1], instability arises when the slope
condition of Vakhitov-Kolokolov is violated. Since,
when $|n(L_+)-n(L_-)|>1$, the relevant theory indicates
that the solution is always unstable, we argue that what
should be happening here is that the violation of the
slope condition of Vakhitov-Kolokolov is not associated
with an instability because it occurs at the same
time as the change of the count of $|n(L_+)-n(L_-)|$
(from $1$ to $0$ or vice versa). 
This is precisely what we numerically illustrate
in Fig. \ref{figLn}. In particular, the left panel concerns
the principally focusing case associated to the first excited
state ``turning branch'' of Fig. \protect\ref{fig4}. There,
given the fact that $n(L_-)=1$ (due to the one zero-crossing
of the configuration itself, which is an eigenfunction
of $L_-$ with a $0$ eigenvalue), we expect the count of 
eigenvalues $n(L_+)$ to change from $2$ to $1$ exactly at
the critical point, precisely as observed in the figure.
On the other hand, the right panel concerning the
ground state branch of Fig. \protect\ref{fig8} for the
principally defocusing case, corresponds to a case with
$n(L_-)=0$. Hence, when the slope condition is violated
(near the linear limit in this case), it has to be true
that $n(L_+)=0$, while after the turning point of the branch,
the slope condition is satisfied and then $n(L_+)=1$, which
again is consonant with the observed stability. These features are clearly
illustrated in the right panel of Fig. \ref{figLn}.
We believe that similar considerations and counts may resolve
apparent paradoxes that seem to be encountered e.g.
in \cite{boris,boris2}.

\section{Two-Mode Dynamics}

So far, we have considered the two-mode reduction as a tool for 
identifying (quite successfully, as shown above) the stationary
states of the underlying problem. However, here we illustrate how
the same tool can be used to understand the system dynamics.
The two principal dynamical features of the (symmetric) double well system
involve the oscillations of matter between the two wells 
(for low population asymmetries between the wells) and the
nonlinearly induced self-trapping regime (for high population 
asymmetries between the wells); see e.g., \cite{smerzi,Bergeman_2mode}.
For this reason, although our setting here is inherently asymmetric
(due to the nature of $\Gamma(x)$), it is of interest to develop 
a variant of the two-mode approximation that accounts for the
population imbalance between the wells. 

\begin{figure}[tbhp!]
     \centering
           \includegraphics[width=.2\textwidth,angle=270]{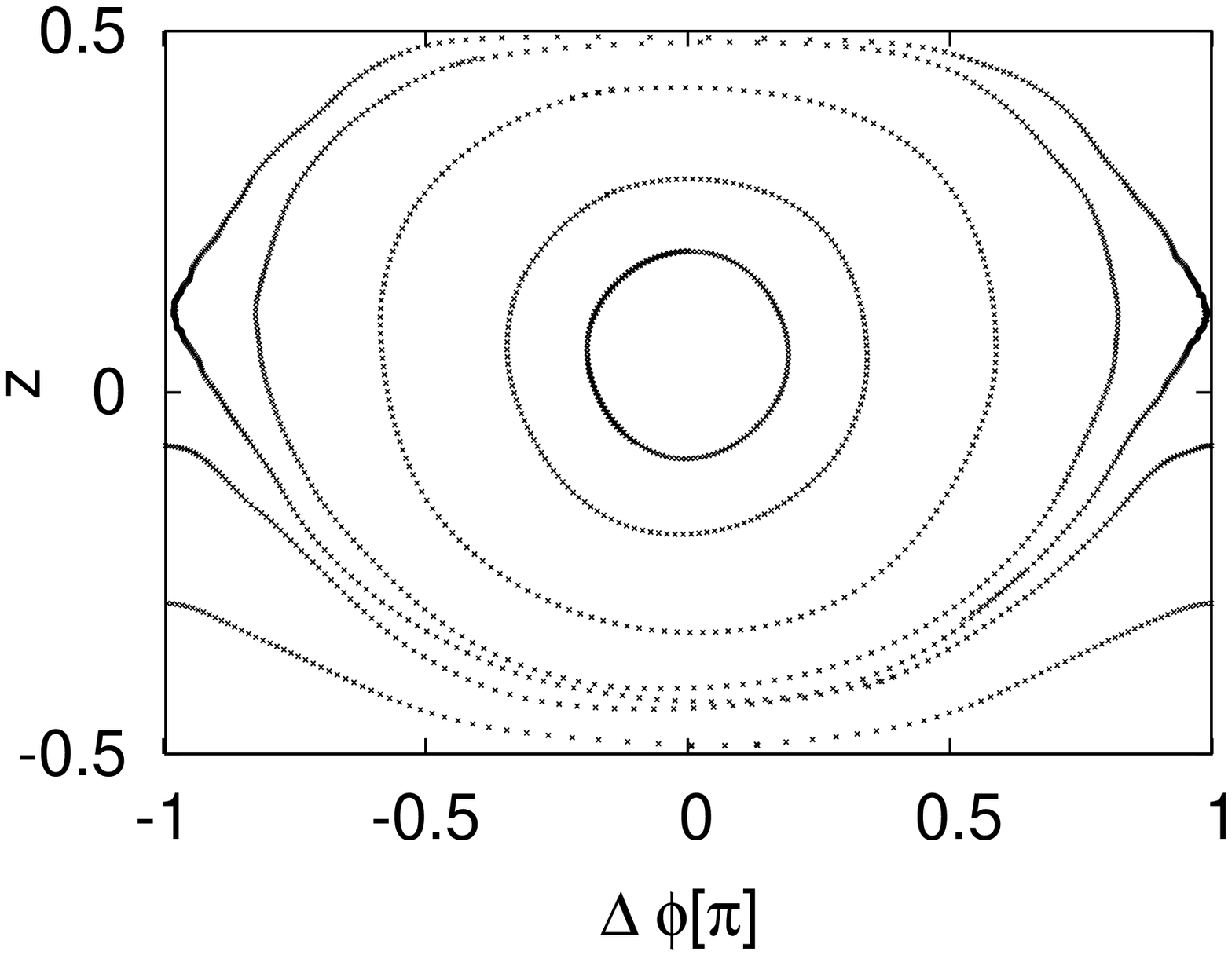}
           \includegraphics[width=.2\textwidth,angle=270]{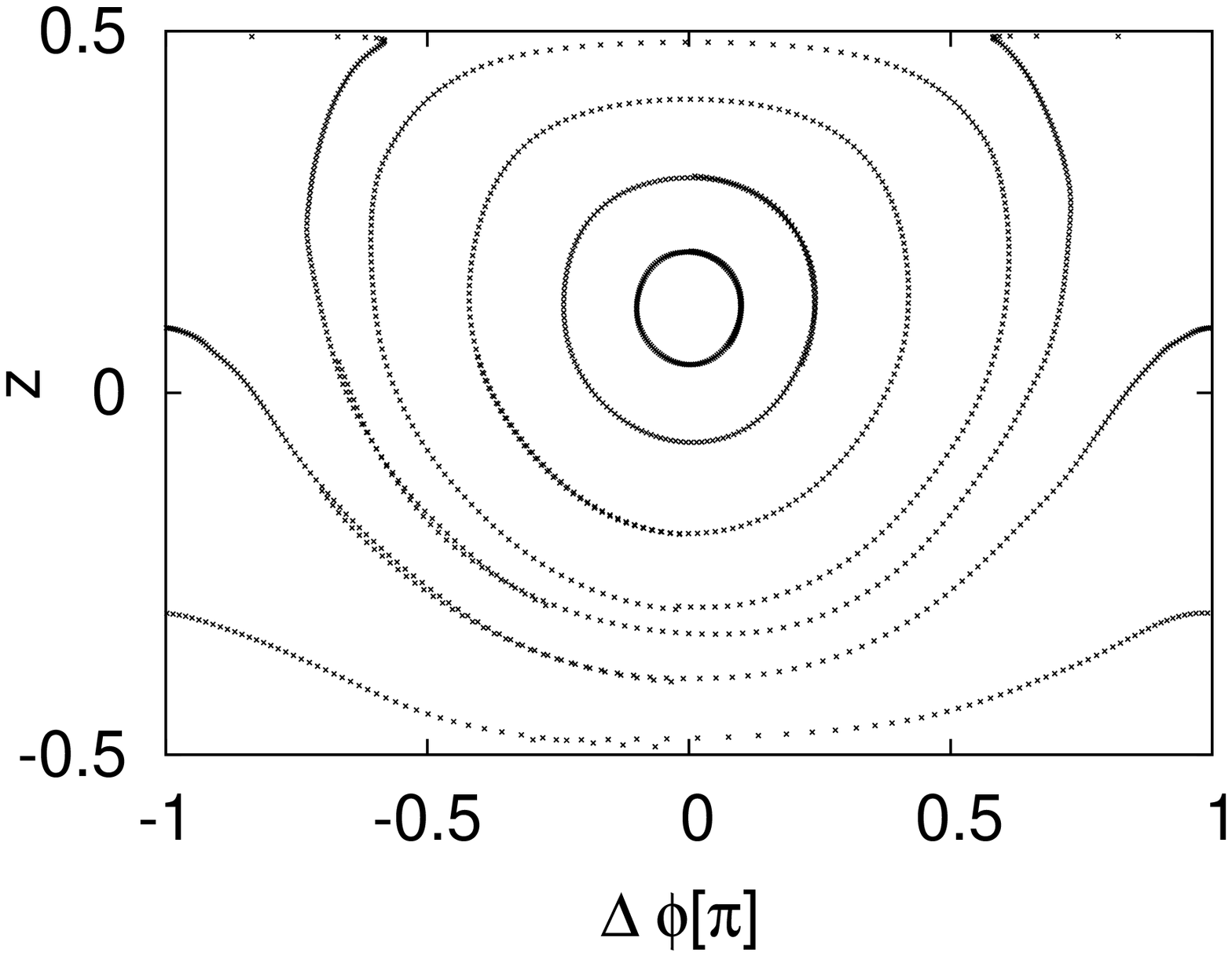}\\  
           \includegraphics[width=.2\textwidth,angle=270]{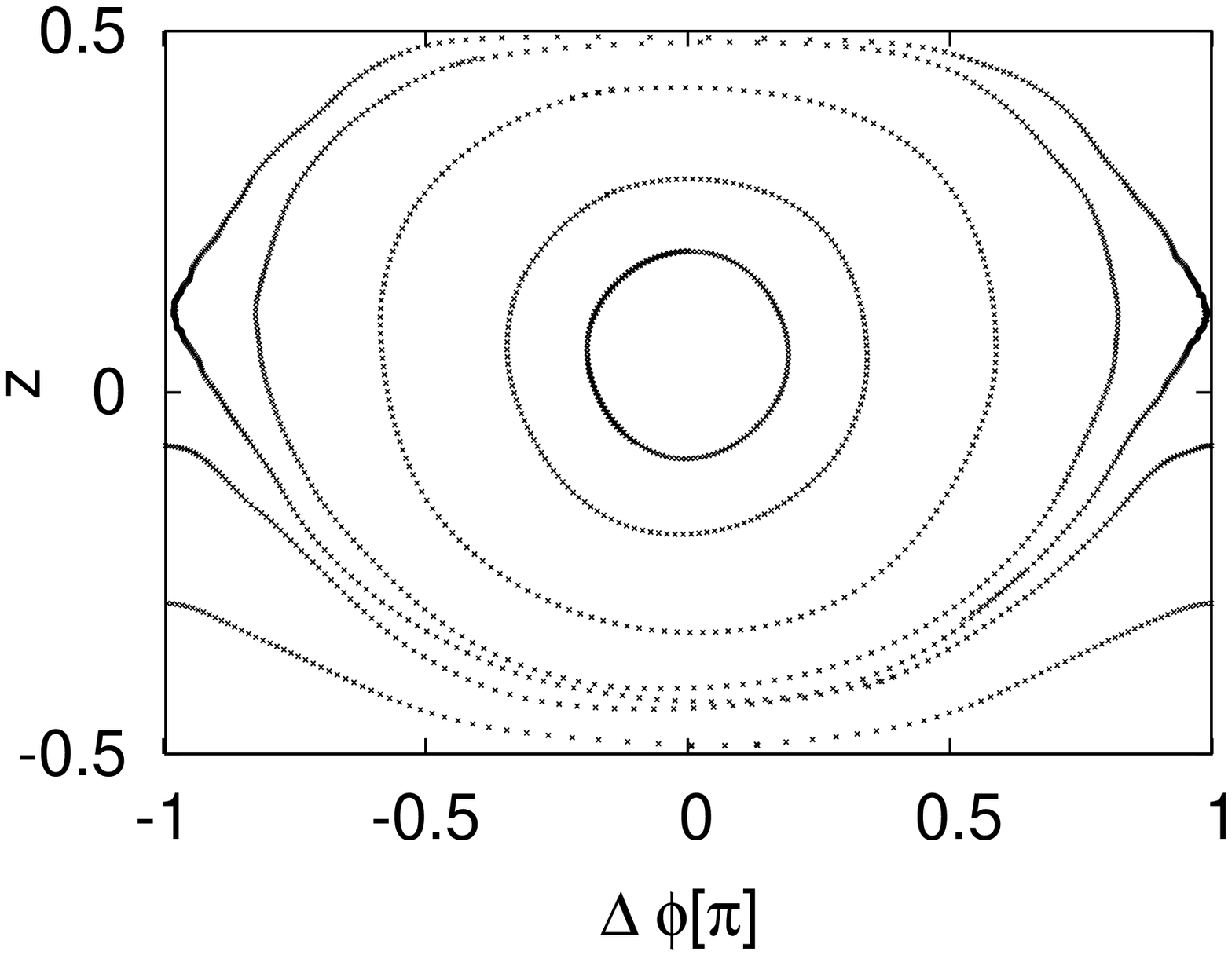}
           \includegraphics[width=.2\textwidth,angle=270]{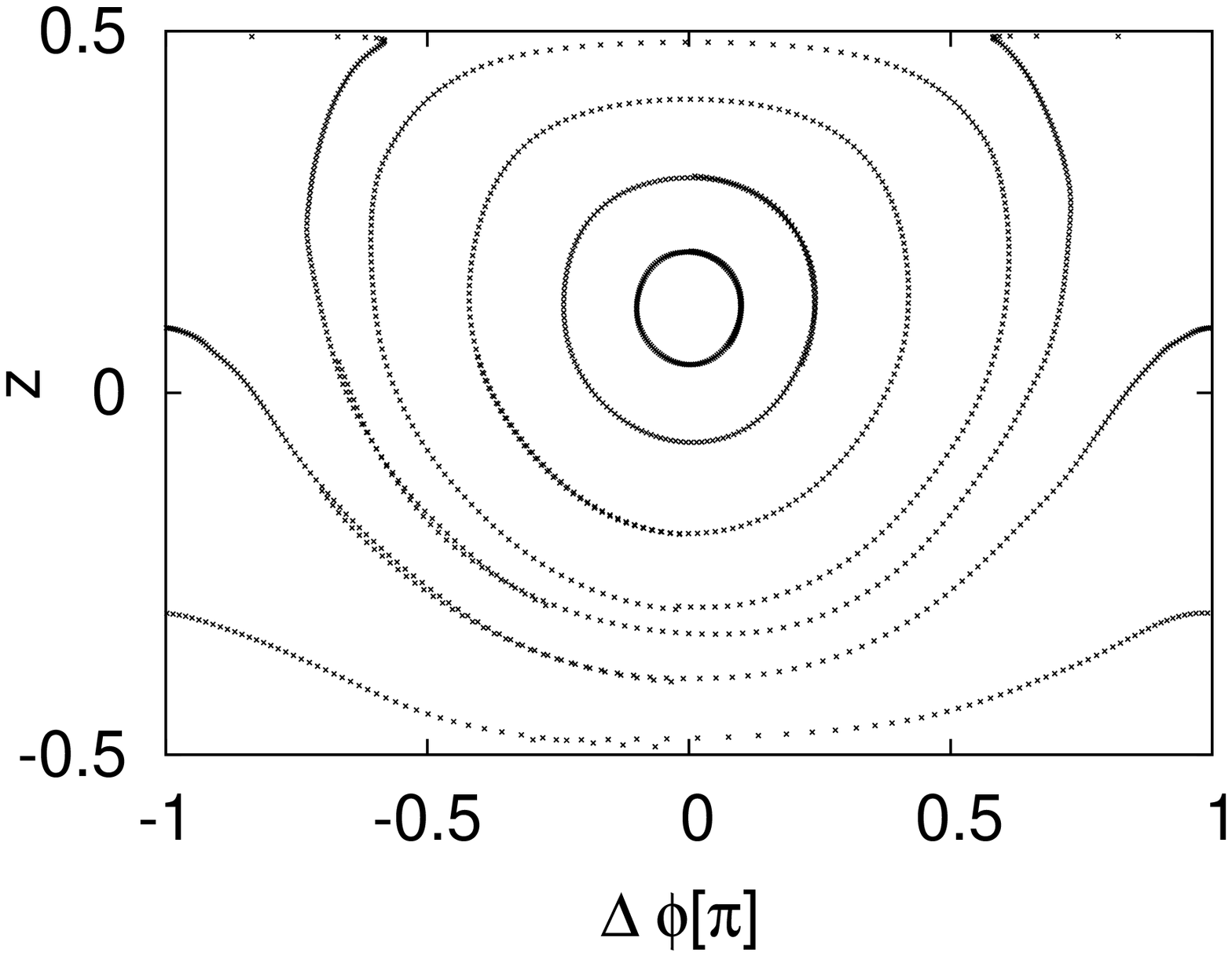}\\  
            \caption{Phase diagrams for 
$\epsilon=0.25$ (left column) and $\epsilon=0.5$ (right column) and $N=0.5$.
The results from the ordinary differential equations (\ref{ode1})-(\ref{ode2})
are shown in the top row, while the corresponding results from the partial
differential equation are shown in the bottom row.}
      \label{figadd1}
\end{figure}

\begin{figure}[tbhp!]
     \centering
\includegraphics[width=.4\textwidth]{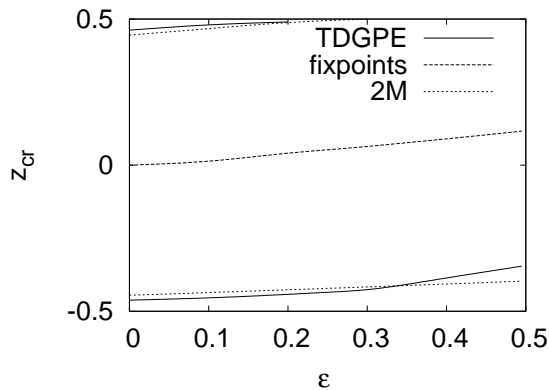}
\caption{The critical points $z_{cr}$ of population imbalance
above which there exists self-trapping are shown in the time-dependent
Gross-Pitaevskii equation (TDGPE) and in the two-mode approximation (2M) for 
$N=0.5$.
Notice that in this setting, the population imbalance becomes nonzero
when $\epsilon \neq 0$ due to the asymmetry (as shown by the dashed line).}
      \label{figadd2}
\end{figure}

In order to formulate the problem based on the populations of 
the left and right well, one can reformulate our two-mode 
decomposition as
\begin{equation}
 u(x,t)=c_R(t)\frac{1}{\sqrt{2}}(u_0(x)+u_1(x))+c_L(t)\frac{1}{\sqrt{2}}(u_0(x)-u_1(x)), \label{Ansatz twomode}
\end{equation}
where it is clear that $c_L$ and $c_R$ are connected with $c_0$ and $c_1$
through a simple linear transformation. If we then decompose
 $c_i=\rho_ie^{\imath \phi_i}$ $i=R,L$ and define 
the phase difference $\Delta \phi = \phi_L-\phi_R$, and the population
imbalance  $z=\rho_R^2-\rho_L^2$ (recall that $N=\rho_L^2+\rho_R^2$),
one can then project the equation to $(u_0+u_1)$, as well as
to $(u_0-u_1)$ and eventually obtain the dynamical equations for the
conjugate variables $z$ and $\Delta \phi$. These read:
\begin{eqnarray}
\dot{\Delta \phi} &=& -\frac{z}{4}\bigl(A_0-10 B+ A_1\bigr)+N\bigl(D_0+D_1\bigr)
-\cos(\Delta\phi)\frac{z}{\sqrt{N^2-z^2}}\Bigl(\mu_0-\mu_1+\frac{N}{2}\bigl(A_0-A_1\bigr) \Bigr)
\nonumber\\
&&
+\cos(\Delta\phi)\frac{N^2-2z^2}{\sqrt{N^2-z^2}}\bigl(D_0-D_1\bigr)
-\cos(2 \Delta \phi)\frac{z}{4}\bigl(A_0-2B+A_1\bigr)
\label{ode1}
\\
\dot{z}&=&\sqrt{N^2-z^2} \sin(\Delta \phi)\Bigl(\mu_0-\mu_1+\frac{N}{2}\bigl(A_0-A_1\bigr)+z\bigl(B_0-B_1\bigr)\Bigr)
+\frac{N^2-z^2}{4} \sin(2 \Delta \phi)\bigl(A_0-2B+A_1\bigr)
\label{ode2}
\end{eqnarray}
It is interesting to examine the dynamical evolution of these
equations (their stationary states are identical to the ones
identified above), and to compare their dynamics with the
corresponding PDE dynamics. In the latter, one can also
make similar projections to $(u_0+u_1)$, as well as
to $(u_0-u_1)$ and define accordingly $c_{L,R}$, as well
as thereafter $z$ and $\Delta \phi$. A comparison of the 
$(z,\Delta \phi)$ phase space for different $\epsilon$ of such canonically 
conjugate variables can be found in Fig. \ref{figadd1}.
It can be clearly observed that for non-zero $\epsilon$ the reflection symmetry around
 $z=0$ is broken. For increasing $\epsilon$ the population imbalance of the stationary states shifts
towards larger values of $z$ (see also Fig. \ref{figadd2}), denoting a larger occupation of the right well compared to the left one.
This can be understood by taking into account that we consider the case of repulsive interaction.
There, an increase of $\epsilon$ implies a reduction of interaction in the right well (see Fig. \ref{fig5}).
So, pictorially speaking, the atoms feel less repulsion in the right well than in the left one, leading to 
the observed population imbalance. For small populational asymmetries (around
the nonzero stationary ones), the system executes inter-well matter-transfer oscillations. On the other
hand, beyond a critical asymmetry in the initial populations, 
we enter the self-trapping regime, as is shown both in the partial
differential equation of the full dynamics and in the above reduced
two-mode description of Eqs. (\ref{ode1})-(\ref{ode2}). For non-zero $\epsilon$ the value of this critical points taken at $\delta \phi=0$
depend also on their sign, in contrary to the symmetric case for $\epsilon=0$. For small values of $\epsilon$ there are two critical points, which increase with increasing $\epsilon$. Eventually, the positive one ceases to exist as one can see by inspecting Fig. \ref{figadd1} for $\epsilon=0.5$. 

A detailed discussion of the dynamical comparison of the two-mode
approximation with the corresponding results of the full system can be found in 
\cite{Bergeman_2mode}. There, it is highlighted how the effect of a
reduced barrier height of the trapping potential or increased interactions can deteriorate the
dynamical effectiveness of the two-mode reduction.
 
Here we explore the role of the collisional inhomogeneity in 
affecting the accuracy of this reduction, by illustrating 
a prototypical diagnostic, namely the critical population imbalance
threshold beyond which self-trapping occurs, as this is identified
in the full dynamical equation and as it is obtained within the two-mode
reduction. This is shown in Fig. \ref{figadd2}. It can be seen that
as $\epsilon$ increases, both critical points (the positive and the negative one) increase.
The positive one vanishes when it is equal to the total number of atoms. By comparing the results obtained by solving the PDE
with the results of the ODEs one observes that the two-mode approximation becomes
less accurate, as $\epsilon$ increases, in capturing the corresponding threshold, a feature
that we have also seen regarding stationary state properties; see e.g.
Fig. \ref{fig7}.

\section{Conclusions and outlook}

In this paper, we 
studied in detail the nature of the most fundamental 
matter-waves  
(emanating from the linear limit of the problem) that emerge from a 
quasi-1D collisionally inhomogeneous Bose-Einstein condensate 
trapped in a 
double-well potential. 
We specifically considered a setup which 
features distinct scattering lengths 
between the two wells, including also the case 
where the nonlinearity coefficient of the pertinent Gross-Pitaevskii equation has different signs 
in the the two wells. We observed that the
relevant phenomenology is different in the considered collisionally
inhomogeneous environment in comparison to the collisionally homogeneous
case studied previously. In particular, even for 
weak inhomogeneities, the asymmetry (that the spatial dependence of the nonlinearity introduces)
induces a modification of the bifurcation picture of the double-well potential 
and a change in the nature of the symmetry-breaking bifurcation 
from a pitchfork to a saddle-node; this is reminiscent of similar 
modifications to the bifurcation picture due to asymmetries
of the linear potential \cite{theo}. 
On the other hand, it was found that as the
strength of the inhomogeneity is increased, even this saddle-node
``recollection'' of the symmetry-breaking bifurcation eventually disappears
(the corresponding critical point is pushed to infinity) and only
one nonlinear branch persists that corresponds to each state of the problem's 
linear limit. Interestingly also, as the inhomogeneity acquires opposite 
sign values of the scattering length, the monotonicity of the number of
atoms' dependence on the chemical potential may change (and may even be
different between the two branches), although they do maintain their
stability (which presents an interesting deviation --in this spatially
inhomogeneous case-- from the well-known Vakhitov-Kolokolov criterion
that we rationalized in detail above).
All of this phenomenology, including the detailed bifurcation 
diagram and other 
specific features, such as the critical point of the
saddle-node bifurcation and its dependence on the degree of inhomogeneity, 
are captured remarkably accurately by a Galerkin-type, two-mode 
approximation and 
a resulting simple set of algebraic equations. 
Finally, we also briefly discussed dynamical aspects of the
system including a phase space reduction in the population imbalance-interwell
phase difference space. We have illustrated that despite the ensuing
asymmetry in the phase space (for positive vs. negative population
imbalances), the two-mode reduction can accurately capture the threshold
of transition from Josephson matter-wave oscillations to nonlinearly
induced self-trapping (especially so for weak collisional inhomogeneities).


Our investigation herein 
presents some testable predictions
for the original physical system and its realization in atomic physics 
or, also possibly, in nonlinear optics \cite{zhigang}; 
in particular, as the nonlinearity becomes increasingly different in the two wells, 
the emergence of additional states should be occurring for
increasingly larger number of atoms and eventually, such additional
states (in particular, the stable one among them) should no longer
be observable. It would be of particular interest if such spatial
variations of the nonlinearity could be experimentally realized,
which would allow the direct testing of 
our theoretical predictions.

The work of D.J.F. was partially supported by the Special Research Account of the University of Athens.
P.G.K.\ acknowledges support from NSF-CAREER,
NSF-DMS-0505663 and NSF-DMS-0619492.


\end{document}